\documentclass[oldversion,longauth]{aa}  
\usepackage{graphicx}
\usepackage[authoryear]{natbib}
\bibpunct{(}{)}{;}{a}{}{,} 
\usepackage{txfonts}
\begin{document}
\title{Tracking the impact of environment on the 
galaxy stellar mass function up to $z \sim 1$ 
in the 10k zCOSMOS sample
\thanks{Based on observations obtained at the European Southern 
Observatory(ESO) Very Large Telescope (VLT), Paranal, Chile, as 
part of the Large Program 175.A-0839 (the zCOSMOS Spectroscopic 
Redshift Survey).}
}
\titlerunning{zCOSMOS: Environmental effects on GSMF}
\author{M.~Bolzonella\inst{1} 
\and K.~Kova\v{c}\inst{2}
\and L.~Pozzetti\inst{1}
\and E.~Zucca\inst{1}
\and O.~Cucciati\inst{3}
\and S.~J.~Lilly\inst{2}
\and Y.~Peng\inst{2}
\and A.~Iovino\inst{4}
\and G.~Zamorani\inst{1} 
\and D.~Vergani\inst{1} 
\and L.~A.~M.~Tasca\inst{3,5}
\and F.~Lamareille\inst{6}
\and P.~Oesch\inst{2}
\and K.~Caputi\inst{7,2}
\and P.~Kampczyk\inst{2}
\and S.~Bardelli\inst{1} 
\and C.~Maier\inst{2}
\and U.~Abbas\inst{8}
\and C.~Knobel\inst{2}
\and M.~Scodeggio\inst{5}
\and C.~M.~Carollo\inst{2}
\and T.~Contini\inst{6}
\and J.-P.~Kneib\inst{3}
\and O.~Le~F\`evre\inst{3}
\and V.~Mainieri\inst{9}
\and A.~Renzini\inst{10}
\and A.~Bongiorno\inst{11}
\and G.~Coppa\inst{1,12}
\and S.~de~la~Torre\inst{4,5}
\and L.~de~Ravel\inst{3,7}
\and P.~Franzetti\inst{5}
\and B.~Garilli\inst{5}
\and J.-F.~Le~Borgne\inst{6}
\and V.~Le Brun\inst{3}
\and M.~Mignoli\inst{1}
\and R.~Pell\'o\inst{6}
\and E.~Perez-Montero\inst{6,13}
\and E.~Ricciardelli\inst{10,14}
\and J.~D.~Silverman\inst{2,15}
\and M.~Tanaka\inst{9,15}
\and L.~Tresse\inst{3}
\and D.~Bottini\inst{5}
\and A.~Cappi\inst{1}
\and P.~Cassata\inst{16}
\and A.~Cimatti\inst{12}
\and L.~Guzzo\inst{4}
\and A.~M.~Koekemoer\inst{17}
\and A.~Leauthaud\inst{18}
\and D.~Maccagni\inst{5}
\and C.~Marinoni\inst{19}
\and H.~J.~McCracken\inst{20}
\and P.~Memeo\inst{5}
\and B.~Meneux\inst{11,21}
\and C.~Porciani\inst{22}
\and R.~Scaramella\inst{23}
\and H.~Aussel\inst{24}
\and P.~Capak\inst{25}
\and C.~Halliday\inst{26}
\and O.~Ilbert\inst{3}
\and J.~Kartaltepe\inst{27,28}
\and M.~Salvato\inst{25,29}
\and D.~Sanders\inst{27}
\and C.~Scarlata\inst{25}
\and N.~Scoville\inst{25}
\and Y.~Taniguchi\inst{30}
\and D.~Thompson\inst{31}
}
   \offprints{Micol Bolzonella \\ \email{micol.bolzonella@oabo.inaf.it}}
\institute{INAF - Osservatorio Astronomico di Bologna, via Ranzani 1, 40127 Bologna, Italy 
\and ETH Zurich, Institute of Astronomy, Wolfgang-Pauli-Stra{\ss}e 27, 8093 Zurich, Switzerland 
\and Laboratoire d'Astrophysique de Marseille, Universit\'{e} d'Aix-Marseille, CNRS, 38 rue Fr\'{e}d\'{e}ric Joliot-Curie, 13388 Marseille Cedex 13, France 
\and INAF - Osservatorio Astronomico di Brera, via Brera 28, 20121 Milano, Italy 
\and INAF - IASF Milano, via Bassini 15, 20133 Milano, Italy 
\and Laboratoire d'Astrophysique de Toulouse-Tarbes, Universit\'{e} de Toulouse, CNRS, 14 avenue Edouard Belin, 31400 Toulouse, France 
\and Institute for Astronomy, Royal Observatory, Blackford Hill, Edinburgh, EH9 3HJ, Scotland, United Kingdom 
\and INAF - Osservatorio Astronomico di Torino, 10025 Pino Torinese (TO), Italy 
\and European Southern Observatory, Karl-Schwarzschild-Stra{\ss}e 2, 85748 Garching bei M\"unchen, Germany 
\and INAF - Osservatorio Astronomico di Padova, vicolo dell'Osservatorio 5, 35122 Padova, Italy 
\and Max-Planck-Institut f\"ur Extraterrestrische Physik, Giessenbachstra{\ss}e, 84571 Garching bei M\"unchen, Germany 
\and Dipartimento di Astronomia, Universit\`a di Bologna, via Ranzani 1, 40127 Bologna, Italy 
\and Instituto de Astrofisica de Andalucia, CSIC, Apdo. 3004, 18080 Granada, Spain 
\and Instituto de Astrofisica de Canarias,  V\'{\i}a L\'{a}ctea s/n, 38205, La Laguna, Tenerife, Spain 
\and Institute for the Physics and Mathematics of the Universe (IPMU), University of Tokyo, Kashiwanoha 5-1-5, Kashiwa-shi, Chiba 277-8568, Japan 
\and University of Massachusetts, Amherst, USA 
\and Space Telescope Science Institute, 3700 San Martin Drive, Baltimore, MS 21218, USA 
\and LBNL \& BCCP, University of California, Berkeley, CA 94720, USA 
\and Centre de Physique Th\'eorique, UMR 6207 CNRS, Universit\'e de Provence, Case 907, 13288 Marseille, France 
\and Institut d'Astrophysique de Paris, UMR 7095 CNRS, Universit\'e Pierre et Marie Curie, 98 bis Boulevard Arago, 75014 Paris, France 
\and Universit\"ats-Sternwarte, Scheinerstra{\ss}e 1, 81679 M\"unchen, Germany 
\and Argelander-Institut f\"ur Astronomie, Auf dem H\"ugel 71, 53121 Bonn, Germany 
\and INAF - Osservatorio Astronomico di Roma, via di Frascati 33, 00040 Monteporzio Catone, Italy 
\and DSM/Irfu/Service d'Astrophysique, CEA Saclay, 91191 Gif-sur-Yvette, France 
\and California Institute of Technology, MC 105-24, 1200 East California Boulevard, Pasadena, CA 91125, USA 
\and INAF - Osservatorio Astronomico di Arcetri, Largo E. Fermi 5, 50125 Firenze, Italy 
\and Institute for Astronomy, 2680 Woodlawn Drive, University of Hawaii, Honolulu, HI 96822, USA 
\and National Optical Astronomy Observatory, 950 North Cherry Avenue, Tucson, AZ 85719, USA 
\and Max Planck Institut f\"ur Plasma Physics and Excellence Cluster Universe, Boltzmannstra{\ss}e 2, 85748 Garching bei M\"unchen, Germany 
\and Research Center for Space and Cosmic Evolution, Ehime University, Bunkyo-cho, Matsuyama 790-8577, Japan 
\and Large Binocular Telescope Observatory, University of Arizona, 933 N. Cherry Ave., Tucson, AZ 85721-0065, USA 
}
 
\abstract{
We study the impact of the environment on the evolution of galaxies in
the zCOSMOS 10k sample in the redshift range $0.1 \le z \le 1.0$ over
an area of $\sim 1.5$\,deg$^2$. The considered sample of secure
spectroscopic redshifts contains about $8500$ galaxies, with their
stellar masses estimated by SED fitting of the multiwavelength optical
to near-infrared (NIR) photometry. The evolution of the galaxy stellar
mass function (GSMF) in high and low density regions provides a tool
to study the mass assembly evolution in different environments;
moreover, the contributions to the GSMF from different galaxy types,
as defined by their SEDs and their morphologies, can be quantified.
At redshift $z \sim 1$, the GSMF is only slightly dependent on
environment, but at lower redshifts the shapes of the GSMFs in high-
and low-density environments become extremely different, with high
density regions exhibiting a marked bimodality, not reproducible by a
single Schechter function.  As a result of this analysis, we infer
that galaxy evolution depends on both the stellar mass and the
environment, the latter setting the probability of a galaxy to have a
given mass: all the galaxy properties related to the stellar mass show
a dependence on environment, reflecting the difference observed in the
mass functions.  The shapes of the GSMFs of early- and late-type
galaxies are almost identical for the extremes of the density contrast
we consider, ranging from isolated galaxies to rich group members.
The evolution toward $z=0$ of the transition mass $\mathcal{M}_{\rm
cross}$, i.e., the mass at which the early- and late-type GSMFs match
each other, is more rapid in high density environments, because of a
difference in the evolution of the normalisation of GSMFs compared to
the total one in the considered environment.  The same result is found
by studying the relative contributions of different galaxy types,
implying that there is a more rapid evolution in overdense regions, in
particular for intermediate stellar masses.  The rate of evolution is
different for sets of galaxy types divided on the basis of their SEDs
or their morphologies, tentatively suggesting that the migration from
the blue cloud to the red sequence occurs on a shorter timescale than
the transformation from disc-like morphologies to ellipticals.  Our
analysis suggests that environmental mechanisms of galaxy
transformation start to be more effective at $z<1$.  The comparison of
the observed GSMFs to the same quantities derived from a set of mock
catalogues based on semi-analytical models shows disagreement, in both
low and high density environments: in particular, blue galaxies in
sparse environments are overproduced in the semi-analytical models at
intermediate and high masses, because of a deficit of star formation
suppression, while at $z<0.5$ an excess of red galaxies is present in
dense environments at intermediate and low masses, because of the
overquenching of satellites.}


   \maketitle
%

\section{Introduction}
\label{intro}

The environmental dependence of galaxy properties (colour, star
formation, mass) is well established in the local universe.  At
present many local studies have been carried out to analyse the
influence of environment on colours, luminosities, morphologies,
structural parameters, star formation, and stellar masses: all local
relations can be considered as different faces of the
morphology--density relation shown by \citet{Dressler1980}.

At higher redshifts, this kind of study becomes very difficult,
because the need for large spectroscopic samples of faint galaxies
with a good sampling rate hampers a reliable estimate of the
environment.  Until now, therefore, most of the studies in high
density environments have analysed galaxy clusters or groups and the
more general effect of the environment on field galaxy evolution
remains poorly explored.  The evolution of the galaxy stellar mass
function (GSMF) as a function of the large-scale environment has been
studied in the DEEP2 Galaxy Redshift Survey \citep{Bundy2006},
considering the redshift range $z=0.4 - 1.4$, which limits the
connection between this study and those in the local Universe.

Some remaining open questions are: what is the most important
property leading the evolution of field galaxies? Is the fate of a
galaxy decided once its mass is defined or do some external players
have a role? And, if the environment plays such a role, when does it
start to affect galaxy evolution, and by means of which mechanism?

On the basis of literature results, the full story of galaxies is not
consistently presented.

Most low-redshift studies are based on SDSS data. We try to summarise
the most relevant conclusions, without pretending to be exhaustive.
Some studies assert that the mass is the most important parameter in
galaxy evolution: from the colour bimodality, \citet{Balogh2004}
propose that the properties of star-forming galaxies are mainly
related to their mass and that, to preserve the bimodality without
altering the colours modelled by two Gaussian distributions, the
transformation from late- to early-type galaxies should be rapid in
truncating the star formation and efficient for all luminosities and
environments.  Analogous studies reach similar conclusions:
\citet{Hogg2003} find that blue galaxies show no correlation between
their luminosity/mass and local density at a fixed colour;
\citet{Baldry2006} affirm that the fraction of red galaxies depends on
environment, but not their colour--mass relation.  \citet{Thomas2010}
find that correlations between properties of galaxies in the red
sequence are only driven by galaxy mass.  Furthermore,
\citet{Bosch2008b}, investigating the efficiency of transformation
processes on the SDSS groups catalogue, claim that both the colour and
the concentration of a satellite galaxy are mostly determined by their
stellar mass.

On the other hand, many other studies based on the same SDSS dataset
agree on giving importance, at different levels, to both nature and
nurture in the evolutionary paths of galaxies. In these studies,
environment is not considered a secondary effect and it has an impact
on one or more of the galaxy properties and their relations such as
colour, star formation rate and its spatial variation, structural
parameters, morphology, the presence of active galactic nuclei (AGN),
age, and the timescale of transformation of galaxies
\citep[e.g.][]{Kauffmann2004,Tanaka2004,Bamford2009,Skibba2009,Welikala2009,Cooper2010a,Gavazzi2010,Clemens2006,Bernardi2006,Lee2010,Mateus2007,Mateus2008,Blanton2005,Gomez2003}.  

In addition to considering the importance of the environment on galaxy
evolution, the scale on which the environment is evaluated has been
found to be of huge importance: for instance, another study on the
colour bimodality by \citet{Wilman2010} finds that a correlation of
the colour and the fraction of red galaxies with increasing densities
is seen only on scales smaller than $\sim 1\,h^{-1}$\,Mpc, which is
the characteristic scale on which galaxies are accreted in more
massive dark matter haloes, undergoing the truncation of their star
formation. Other studies dealing with the groups environment support a
similar scenario in which central and satellites galaxies follow
different evolutionary paths, with satellite galaxies falling into
more massive haloes and experiencing a slow transformation because of
the removal of gas by strangulation, resulting in the fading of star
formation \citep{Rogers2010,Weinmann2009,Wel2009,Wel2010,Bosch2008}.

Still at low redshifts, but using 2MASS and LCRS data,
\citet{Balogh2001} distinguished between different environments such
as field, groups, and clusters, finding that luminosity and mass
functions depend on both galaxy type (with steeper functions for
emission line galaxies) and environment (with more massive and
brighter objects being more common in clusters), mainly as a
consequence of the different contributions of passive galaxies.

At higher redshifts, probing the effect of environment on galaxy
evolution becomes more difficult and often this kind of study uses
projected estimators of local density and relies on photometric
redshifts \citep[e.g. ][]{Scoville2007b,Wolf2009}. The main studies
using spectroscopic redshifts analyse data from the two major surveys
of the recent past, DEEP2 \citep{Davis2003} and VVDS
\citep{LeFevre2003}.

\citet{Bundy2006}, using DEEP2 data at $0.4<z<1.4$ and $R_{\rm
AB}<24.1$, estimate the effect of environment on GSMFs: they drew the
conclusion that the quenching of star formation, and then the
transition between the blue cloud and the red sequence, is primarily
internally driven and dependent on mass, even if they detected a
moderate acceleration of the downsizing phenomenon in overdense
regions, where the rise of the quiescent population with cosmic time
appears to be faster, as seen through the evolution of the transition
and quenching masses, $\mathcal{M}_{\rm cross}$ and $\mathcal{M}_{\rm
Q}$.  Using the same dataset complemented by SDSS at low redshifts,
\citet{Cooper2008} studied the connection between the star formation
rate (SFR) and environment, finding hints of a reversal of that
relation from $z\sim 0$, where the mean SFR decreases with local
density, to $z\sim 1$, where a blue population causes an increase in
the mean SFR in overdense regions; nonetheless, the decline of the
global cosmic star formation history (SFH) since $z\sim 1$ seems to be
caused by a gradual gas consumption rather than environment-dependent
processes. A similar result on the reversing relationship
SFR--environment was found by \citet{Elbaz2007}, using GOODS data and
SFR derived from UV and $24\,\mu$m emission.

Using spectroscopic data from the VVDS up to $z\sim 1.5$,
\citet{Cucciati2006} found a steep colour--density relation at
low-$z$, which appeared to fade at higher redshifts. In particular,
they identified differences in colour distributions in low and high
density regimes at low redshifts, whereas at high redshifts the
environment was not found to affect these distributions.  In their
proposed scenario the processes of star formation and gas exhaustion
are accelerated for more luminous objects and high density
environments, leading to a shift with cosmic time in star formation
activity toward fainter galaxies and low density environments.
\citet{Scodeggio2009} studied the stellar mass and colour segregations
in the VVDS at redshifts $z=0.2 - 1.4$, using a density field computed
on scales of $\sim 8$\,Mpc; they found that the colour--density
relation is a mirror of the stellar mass segregation, that in turn is
a consequence of the dark matter halo mass segregation predicted by
hierarchical models.

The effects of environment on both local galaxy properties and their
evolution are still uncertain, keeping the nature versus nurture
debate open. From the aforementioned results, there seems to be some
hint that the galaxy evolutionary path from the blue cloud to the red
sequence depends on environment, but the determination of the
mechanism behind this transformation, its probability of occurring,
its link to both the environment and intrinsic galaxy properties is a
difficult task.  Different physical processes of galaxy transformation
differ in terms of timescales, efficiency and observational
repercussions, such as colour and morphology.  The GSMF is a very
suitable tool for investigating this problem and witnessing the
buildup of galaxies and its dependence on environment.

In this paper, we focus on the effect of environment on field galaxies
using data from COSMOS (Cosmic Evolution Survey) and zCOSMOS; in this
field the most extreme overdense regions such as cluster cores are
almost absent. Parallel and complementary analyses are presented in
\citet{Pozzetti2010}, \citet{Zucca2009}, \citet{Iovino2010},
\citet{Cucciati2010}, \citet{Tasca2009}, \citet{Kovac2010b},
\citet{Vergani2010}, \citet{Moresco2010}, and \citet{Peng2010}.  The
plan of this paper is the following: in Sect.~\ref{data}, we describe
the spectroscopic and photometric datasets and the derived properties
we used to characterise different galaxy populations; in
Sect.~\ref{mf} we derive the GSMFs and in Sect.~\ref{mftype} we
analyse the different contribution of galaxy types to the GSMF in
different environments. We compare our results with similar analyses
in the literature and we discuss the implications for the picture of
galaxy evolution in Sect.~\ref{discuss}.

Throughout the paper we adopted the cosmological parameters
$\Omega_m=0.25$, $\Omega_\Lambda=0.75$, $h_{70}=H_0/(70\,{\rm
km\,s^{-1}\,Mpc^{-1}})$, magnitudes are given in the AB system and
stellar masses are computed assuming the Chabrier initial mass
function \citep{Chabrier2003}.


\section{Data}
\label{data}

The zCOSMOS survey \citep{Lilly2007} is a redshift survey intended to
measure the distances of galaxies and AGNs over the COSMOS field
\citep{Scoville2007}, the largest HST survey carried out to date
with ACS \citep{Koekemoer2007}. The whole field of about $2$\,deg$^2$
was observed from radio to X-ray wavelengths by parallel projects,
involving worldwide teams and observatories. The coexistence of
multiwavelength observations, morphologies, and spectroscopic
redshifts ensures that COSMOS provides a unique opportunity to study
the evolution of galaxies in their large-scale structure context.

\subsection{Spectroscopy}
\label{spectro}

The spectroscopic survey zCOSMOS is currently ongoing and is
subdivided into two different parts: the ``bright'' survey, which
targets $\sim 20\,000$ galaxies, with a pure flux-limited selection
corresponding to $15 \le I_{\rm AB} \le 22.5$, and the ``deep''
survey, whose goal is the measurement of redshifts in the range
$1.4\le z \le 3.0$, within the central $1$\,deg$^2$.

The data used in this paper belong to the so-called 10k sample
\citep{Lilly2009}, consisting of the first $10\,644$ observed objects
of the ``bright'' survey, over an area of $1.402$\,deg$^2$ with a mean
sampling rate of $\sim 33$\%. The final design of the survey aims to
reach a sampling rate of $\sim 60-70$\%, achieved by means of an
eight-pass strategy. The observations have been carried out with
VIMOS@VLT with the red grism at medium resolution $R\sim 600$.  The
data have been reduced with VIPGI \citep{Scodeggio2005} and
spectroscopic redshifts have been visually determined after a first
hint provided by EZ \citep{Garilli2010}\footnote{Both VIPGI and EZ are
public softwares retrievable from {\tt
http://cosmos.iasf-milano.inaf.it/pandora/}}.  The confidence on the
redshift measurements has been represented by means of a flag ranging
from 4, for redshifts assigned without doubts, to 0, for undetermined
redshifts; a subsample of duplicated spectroscopic observations
allowed us to estimate the rate of confirmation of redshift
measurements, being in the range 99.8 -- 70\% depending on the flag
(see \citealt{Lilly2009} for details).  All the redshifts have been
checked by at least two astronomers.  A decimal digit specifies
whether the redshift is in agreement with photometric redshifts
\citep{Feldmann2006} computed from optical and near-infrared (NIR)
photometry using the code ZEBRA \citep[Zurich Extragalactic Bayesian
Redshift Analyzer,][]{Feldmann2008}.  For some objects, the measure
resulted to be hampered by technical reasons (for instance the
spectrum at the edge of the slit); in those cases, a flag $-99$ has
been assigned.  Different flags have been assigned to identify
broad-line AGNs and targets observed by chance in slits.

\subsection{Photometry}
\label{photo}

The photometry used in the following is part of the COSMOS
observations and encompasses optical to NIR wavelengths: $u*$ and
$K_s$ from CFHT, $B_J$, $V_J$, $g^+$, $r^+$, $i^+$, and $z^+$ from
Subaru, and Spitzer IRAC magnitudes at $3.6$, $4.5$, $5.8\,\mu$m.
Details of photometric observations and data reduction are given in
\citet{Capak2007} and \citet{McCracken2010}. The scantiness of
standard stars in the photometric observations and the uncertainty in the
knowledge of the filter responses result in an uncertain calibration
of zero-points.  To avoid this inconvenience, we optimised the
photometry by applying offsets to the observed magnitudes: we computed
these photometric shifts for each band minimising the differences
between observed magnitudes and reference ones computed from a set of
spectral energy distributions (hereafter SEDs). We adopted an approach
similar to \citet[see their Table 13]{Capak2007}, but considering the
same set of SEDs we used to compute stellar masses detailed in
Sect.~\ref{stellarmasses}, obtaining in general very similar offsets
for all the filters.

\subsection{Stellar masses}
\label{stellarmasses}

Stellar masses were evaluated by means of a SED fitting technique,
using the code \emph{Hyperzmass}, a modified version of the
photometric redshift code \emph{Hyperz} \citep{Bolzonella2000}.
\citet{Marchesini2009} analysed the effect of random and systematic
uncertainties in the stellar mass estimates on the GSMF, considering
the influence of metallicity, extinction law, stellar population
synthesis model, and initial mass function (IMF). On the other hand,
\citet{Conroy2009} analysed the impact of the choice of the reference
SEDs on the output parameters of the stellar population synthesis.
Here we describe the approach and the tests we performed on our data.

We used different libraries of SEDs, derived from different models of
stellar population synthesis: (1)~the well-known \citet[hereafter
BC03]{Bruzual2003} library, (2)~\citet[hereafter M05]{Maraston2005}
and (3)~\citet[hereafter CB07]{Charlot2010}. The main difference
between the three libraries is the treatment of thermally pulsing
asymptotic giant branch (TP-AGB) stars. M05 models include the TP-AGB
phase, calibrated with local stellar populations. This stellar phase
is the dominant source of bolometric and NIR energy for a simple
stellar population in the age range $0.2$ to $2$\,Gyr. Summing up the
effects of both overshooting and TP-AGB, the M05 models are brighter
and redder than the BC03 models for ages between $\sim 0.2$ and $\sim
2$\,Gyr \citep{Maraston2006}. The use of the M05 models leads to the
derivation of lower ages and stellar masses for galaxies in which the
TP-AGB stars are contributing significantly to the observed SED (i.e.,
ages of the order of $\sim 1$\,Gyr).  At older ages, the M05 models
are instead bluer. CB07 is the first release of the new version of the
Charlot \& Bruzual library, which is not yet public. CB07 models
include the prescription of \citet{Marigo2007} for the TP-AGB
evolution of low and intermediate-mass stars. As for the M05 models,
this assumption produces significantly redder NIR colors, hence
younger ages and lower masses for young and intermediate-age stellar
populations.  A brief description of the effect on GSMFs of different
choices of template SEDs can be found in the companion paper by
\citet{Pozzetti2010}.

All the considered libraries provide a simple stellar population (SSP)
and its evolution in many age steps for a fixed metallicity and a
given IMF; it is possible from the SSP models to derive the composite
stellar populations that can reproduce the different types of observed
galaxies, imposing a star formation history (SFH). We compiled 10
exponentially declining SFHs with $e$-folding times ranging from $0.1$
to $30$\,Gyr plus a model with constant star formation.  Smooth SFHs
are a simplistic representation of the complex SFHs galaxies have
experienced. In \citet{Pozzetti2007}, using VVDS data, we also
computed stellar masses using SEDs with random secondary bursts
superimposed on smooth SFHs, finding average differences well within
the statistical uncertainties for most of the sample. However,
repeating the comparison with the zCOSMOS 10k sample, we estimated
that about $15$\% of the sample has $\log {\mathcal M}_{\rm complex} /
{\mathcal M}_{\rm smooth} \ga 0.35$\,dex \citep[see
also][]{Pozzetti2010}.  Most of these galaxies are characterised by a
significant fraction of stellar mass ($\sim 5$ -- $15$\%) produced in
a secondary burst in the past Gyr and an age of the underlying
smoothly evolving population a few Gyr older than the age obtained by
fitting SEDs with only smooth SFHs.  We verified that these
differences in the stellar mass estimate produce negligible effects on
the final GSMF and therefore the results are not affected by the
choice of the SEDs.

The IMF is another important parameter: different choices on the IMF
produce different estimates of stellar mass, but these differences can
be statistically recovered. The most widely used IMFs are those of
Salpeter \citep{Salpeter1955}, Kroupa \citep{Kroupa2001}, and Chabrier
\citep{Chabrier2003}.  The statistical differences in stellar masses
are given by $\log {\mathcal M}_{\rm Salp}\simeq \log {\mathcal
M}_{\rm Chab} +0.23$ and $\log {\mathcal M}_{\rm Chab}\simeq
\log{\mathcal M}_{\rm Krou} -0.04$.  Using the zCOSMOS and a mock
photometric catalogue, we checked how the other parameters of the SED
fitting, i.e. the age and the amount of reddening, vary when the SEDs
are compiled using Chabrier and Salpeter IMFs: we found that these
parameters are very similar for the two best-fit SEDs, with negligible
offset and very small dispersion. In the following, stellar masses are
computed assuming the Chabrier IMF.

In stellar population synthesis models, the metallicity can either
evolve with time or remain fixed. In BC03, the included software does
not allow us to build SEDs with evolving metallicity, although $6$
different values of $Z$ are available.  To evaluate the effect of
metallicity on stellar masses and GSMFs, we verified in simulated and
real catalogues that the inclusion of different values of $Z$ does not
introduce a significant bias, the differences on the best-fit stellar
masses being $\la 0.1$\,dex.  Using the available values of $Z$ does
not lead to a substantial improvement in the quality of the best-fits,
at the cost of the introduction of an additional parameter.
We therefore adopted a fixed and solar metallicity.

Dust extinction was modelled using the Calzetti's law
\citep{Calzetti2000}, with values ranging from $0$ to $3$ magnitudes
of extinction in $V$ band.

The $\chi^2$ minimisation comparing observed and template fluxes at a
fixed redshift $z=z_{\rm spec}$ provides the best-fit SED, with which
are associated a number of physical parameters, such as age,
reddening, instantaneous star formation, and stellar mass.  We note
that the meaning of stellar mass throughout this paper is not the
integral of the star formation, because from that value we would have
to exclude the return fraction, i.e., the fraction of gas processed by
stars and returned to the interstellar medium during their evolution.

Tests on simulated catalogues considering the effect on stellar mass
estimates of different choices of reddening law, SFHs, metallicities,
and SED libraries show a typical dispersion of the order of
$\sigma_{\log\mathcal{M}}\simeq 0.20$.  Even a simpler technique such
as that used by \citet{Maier2009} and derived from Eq.~1 of
\citet{Lin2007}, produces a scatter not larger than $\sim 0.16$\,dex,
although with some slight trend as a function of stellar mass and
redshift.  These tests show that stellar mass is a rather stable
parameter in SED fitting when dealing with a set of data spanning a
wide wavelength range extending to NIR.

Since the fluxes provided by the available libraries at IR wavelengths
have been extrapolated, the choice of filters used in determining
best-fit solutions is limited to $2.5\,\mu$m rest-frame for M05 models
(at longer wavelengths, these models use the Rayleigh-Jeans tail
extrapolation) and to $5\,\mu$m rest-frame for BC03 and CB07 models,
since at longer wavelengths the dust re-emission can contribute to the
flux budget.

A problem arising when dealing with a very large number of template
SEDs is to avoid non-physical best-fits. We applied two priors (the same
used in \citealp{Pozzetti2007}, and proposed by \citealp{Fontana2004}
and \citealp{Kauffmann2003}) to avoid such a problem. In particular,
we excluded best-fit SEDs not fulfilling the following requirements:
(1)~$A_V \le 0.6$ if age$/\tau \ge 4$ (i.e., old galaxies must have a
moderate dust extinction); (2)~star formation must start at $z > 1$ if
$\tau < 0.6$\,Gyr (to obtain a better estimate of the ages of
early-type galaxies typically fitted by these low-$\tau$ models).
Moreover, we tested by means of simulations that imposing a minimum
best fit age of $0.09$\,Gyr reduces potential degeneracies and
improves the reliability of the stellar mass estimate. The maximum
allowed age is the age of the Universe at $z_{\rm spec}$.

As mentioned in Sect.~\ref{photo}, the first SED fitting run over the
brightest galaxies and most secure galaxy redshifts has been performed
to compute the photometric offsets. We checked that additional
iterations of the SED fitting and offset estimation do not
significantly improve the $\chi^2$ statistics.

To ease the comparison with literature results, in the following we
present GSMFs obtained adopting the BC03 stellar masses. However, the
qualitative trends are the same for any choice of stellar population
synthesis model.

\subsection{Environment}
\label{env}

The density field was derived for the 10k spectroscopic sample using
different estimators combined with the ZADE \citep[Zurich Adaptive
Density Estimator,][]{Kovac2010a} algorithm.  Some of the existing
studies rely heavily on photometric redshifts and projected densities
computed in wide redshift slices, possibly diluting the signal from
overdense regions.  \citet{Cooper2005} found that photometric
redshifts with accuracies of $\sigma_z \ga 0.02$ hamper the
computation of the density field on small scales.  An important added
value of COSMOS is the availability of spectroscopic redshifts
obtained with a good sampling rate, making feasible an accurate
estimate of the environment, with high resolution also on the radial
direction.

To this aim, we used spectroscopic redshifts to delineate a skeleton
of galaxy structures, and we incorporated a statistical treatment of
the likelihood function of photometric redshifts computed with ZEBRA.
This approach allows us to probe a wide range of environments, thanks
to the precision of spectroscopic redshifts, and to reduce the Poisson
noise, thanks to the inclusion of fractional contributions belonging
to objects with photometric redshifts, estimated from their
probability function. Results have been extensively and carefully
tested on mock catalogues from the Millennium simulation
\citep{Kitzbichler2007}. The reconstruction of overdensities
$1+\delta$ has been explored using different tracer galaxies,
different spatial filters, and different weights (e.g., luminosity or
stellar mass) assigned to each galaxy. The density contrast $\delta$
is defined as $(\rho-\bar{\rho})/\bar{\rho}$, where $\rho$ is the
density as a function of RA, DEC, and $z$ and $\bar{\rho}$ is the mean
density measured at the same redshift.  In principle, a fully
realistic physical representation of the environment should involve
the mass of the dark matter haloes in which the galaxies are
embedded. This mass is clearly not directly accessible to
observations, hence an affordable surrogate to weight the number
density field is given by the stellar masses of the surrounding
galaxies. This is a proxy of the overall density field, since galaxies
are biased tracers of the underlying matter distribution.  The choice
of a fixed selection band results in different populations
preferentially sampled at different redshifts, weighting with stellar
mass should also mitigate this issue.  As expected, mass-weighted
overdensities have an increased dynamical range, in particular at the
highest densities.  As we see in Sect.~\ref{mfenv}, this procedure,
although physically motivated, can introduce some spurious signal,
mainly induced by the mass of the galaxy around which the overdensity
is computed.

Another estimate of the high density environments in which galaxies
reside can be obtained by selecting optical groups, as described in
\citet{Knobel2009}, or X-ray ones (\citealt{Finoguenov2007};
Finoguenov et al. \citeyear{Finoguenov2010}); low density environments
can be tracked by isolated galaxies defined using their Voronoi
volumes, as in \citet{Iovino2010}. The two determinations of the
environment are in fairly good agreement, considering the differences
of the involved scales, with most galaxies being members of groups
residing in the most overdense regions (see also Sect.~\ref{defD1D4}).

In the following, we use as reference the $5$th nearest neighbour
estimator (hereafter 5NN) of the density field, which represents a
good compromise between the smallest accessible scales and the
reliability of the overdensity values. In this approach, tracer
galaxies, selected to be brighter than absolute magnitudes
$M_B=-20.5-z$ or $M_B=-19.3-z$, are considered within an interval $\pm
1000$\,km\,s$^{-1}$ centred on the central galaxy and counted, after
distance sorting, until their number becomes larger than $5$,
considering also the fractional contribution from objects with
photometric redshifts.  Photometric redshifts are not crucial to the
estimate of the density field, but they mainly contribute to reduce
the Poisson noise and improve the agreement with the ``true'' density
field, as has been proven by testing the method on simulated samples.
Overdensities are then computed at the position of each galaxy in the
spectroscopic sample, considering also the contribution to the number
or mass density of the galaxy itself.  We checked that the same
qualitative trends of the GSMFs analysed in the following are present
also when considering other estimators.

\subsection{Galaxy type classification}
\label{class}

Galaxy types can be classified in a multitude of ways, using their
rest-frame colours, their SEDs, their spectroscopic features, their
structural parameters and their morphologies, all of them derivable
with different methods. Different classifications map different
physical properties.  For instance, the rest-frame colour $U-B$ and
the galaxy SED are used as a proxy of the star formation activity and
history, the morphology is an indicator of the dynamical state, and
the two are partially independent \citep{Mignoli2009}.

Even if COSMOS offers a wide range of methods to group galaxies, we
chose to use only two types of classification: photometric and
morphological.

The photometric type is defined by SED fitting to the optical
magnitudes, assuming as reference the same templates used by
\citet{Ilbert2006}: the four locally observed CWW \citep{Coleman1980}
and two starburst SEDs from \citet{Kinney1996}, extrapolated at UV and
mid-IR wavelengths. These six templates are then interpolated to
obtain 62 SEDs and optimised with VVDS spectroscopic data.  The SED
fitting, a $\chi^2$ minimisation performed with the code ALF
\citep{Ilbert2005,Zucca2006,Zucca2009}, provides as output the
best-fit solution.  Galaxies are then classified into two types,
closely corresponding to colours of ellipticals up to early spirals
(type 1, hereafter T1) and later types up to irregular and starburst
galaxies (type 2, hereafter T2) to explore in a simple way the
evolution of the early- and late-type bimodality.

We adopted the morphological classification presented in
\citet{Scarlata2007}: the availability of deep F814-band HST ACS
images over the whole COSMOS field \citep{Koekemoer2007} allows a good
determination of the structural parameters on which the morphology
derived with the software ZEST \citep[Zurich Estimator of Structural
Types,][]{Scarlata2007} is based. The method is a PCA analysis using
estimates of asymmetry, concentration, Gini coefficient, $M_{20}$ (the
second order moment of the $20$\% brightest pixels), and ellipticity.
The morphological classes are the following: early-type (type 1), disk
(type 2, with an associated sub-classification ranging from $0$ to $3$
representing the ``bulgeness'', derived from the $n$ S\'ersic indices,
\citealt{Sargent2007}), and irregular galaxies (type 3). Adopting the
same line of reasoning used for the photometric types, we grouped
morphologically classified galaxies into two broad classes, with
early-type including classes $1$ and $2.0$, i.e., ellipticals and
bulge-dominated galaxies.


\section{Mass functions}
\label{mf}

\subsection{The sample}
\label{sample}

Not all the spectroscopic redshifts have the same level of
reliability, as explained in Sect.~\ref{spectro}. The sample we used
includes only the galaxies with flags corresponding to most secure
redshifts, i.e., starting from flag $=1$ in case of agreement with
photometric redshifts.  In detail, we excluded from our sample broad
line AGNs ($\sim 1.8$\% of the statistical sample), stars ($\sim
5.9$\%), objects with fewer than five detected magnitudes available to
compute the SED fitting ($\sim 1.7$\%) and objects for which the
ground photometry can be affected by blending of more sources, as
derived from the number of ACS sources brighter than $I = 22.5$ within
$0.6''$ ($\sim 0.5$\%).  The final sample contains $8450$ galaxies
with redshifts between $0.01$ and $2$ and $7936$ in the redshift range
where the following analysis is carried out, $z=0.1-1$.  For this
sample, the global reliability of spectroscopic redshifts is $96$\%,
as estimated from the mix of flags and the associated verification
rates reported in \citet{Lilly2009}.

\subsection{Statistical weights}
\label{weights}

We took into account that the observed galaxies are only a fraction of
the total number of possible available targets with the same
properties by applying statistical weights to each observed object
\citep{Zucca1994,Ilbert2005}. We computed the weight $w_i$ for each
galaxy in our sample as the product of two factors connected to the
target sampling rate (TSR) and to the spectroscopic success rate
(SSR). Here we outline the basic principles on which the computation
is based, referring the reader to \citet{Zucca2009} for further
details.

The TSR is the fraction of sources observed in the spectroscopic
survey compared to the total number of objects in the parent
photometric catalogue from which they are randomly extracted.  In the
case of zCOSMOS, the VMMPS tool for mask preparation
\citep{Bottini2005} has been set in such a way that the objects have
been randomly selected without any bias. A different treatment has
been granted to compulsory targets, i.e., objects with forced slit
positioning: they have a much higher TSR ($\sim 87$\%) than the
``random'' sample ($\sim 36$\%). The associated weight is $w_i^{\rm
TSR}=1/{\rm TSR}$.

The SSR represents the fraction of observed sources with a
successfully measured redshift: it is a function of apparent
magnitude, being linked to the signal-to-noise ratio of the spectrum,
and it ranges from $97.5$\% to $82$\% for the brightest and faintest
galaxies, respectively. The weight derived from the SSR is $w_i^{\rm
SSR}=1/{\rm SSR}$.

The SSR is not only a function of magnitude, but also of redshift,
since the spectral features on which the redshift measurement relies
can enter or go out of the observed wavelength window
\citep{Lilly2007}.  Therefore, the redshift distribution of the
measured redshifts can be different from the real one; it is possible
to take into account our lack of knowledge of the failed measurements
by using photometric redshifts. Hence, we used the \citet{Ilbert2009}
release of $z_{\rm phot}$ and computed the SSR in $\Delta z=0.2$
redshift bins.  We had also to consider that the characteristic
emission or absorption lines are different for different galaxy types,
as shown in \citet{Lilly2009}.  We computed the SSR in each redshift
bin by separating red and blue galaxies, selected on the basis of
their rest-frame $U-V$ colour.  The so-called secondary targets, i.e.,
objects in the parent catalogue, imaged in the slit by chance, were
considered separately: they are characterised by a lower SSR because
they are often located at the spectrum edge or observed only at their
outskirts.  We computed and assigned the final weights $w_i=w_i^{\rm
TSR}\times w_i^{\rm SSR}$ considering all the described dependencies.

\subsection{Mass function methods}
\label{mf_method}

To compute the GSMFs, we adopted the usual non-parametric method
$1/V_{\rm max}$ \citep{Avni1980}, from which we derived the best-fit
Schechter function \citep{Schechter1976}. The observability limits
inside each redshift bin, $z_{\rm min}$ and $z_{max}$, were
computed for each galaxy from its best-fit SED.

As in \citet{Pozzetti2010}, we estimated the parametric fit of the
GSMFs with both a single Schechter function, as in most published
results, and the sum of two Schechter functions, which appears to
provide a more accurate fit to the data at least in the lowest
redshift bins.  We adopted the formalism introduced by
\citet{Baldry2004,Baldry2006} using a single $\mathcal{M}^*$ to limit
the number of free parameters
\begin{eqnarray}
\phi(\mathcal{M}){\rm d}\mathcal{M} & = & \phi^*_1\left(\frac{\mathcal{M}}{\mathcal{M}^*}\right)^{\alpha_1} \exp\left(-\frac{\mathcal{M}}{\mathcal{M}^*}\right) {\rm d}\frac{\mathcal{M}}{\mathcal{M}^*} +\nonumber \\
&& + \phi^*_2\left(\frac{\mathcal{M}}{\mathcal{M}^*}\right)^{\alpha_2} \exp\left(-\frac{\mathcal{M}}{\mathcal{M}^*}\right) {\rm d}\frac{\mathcal{M}}{\mathcal{M}^*}\,.
\end{eqnarray}
Until now the need to model a faint-end upturn has been studied in
luminosity function (LF) studies, both in the field
\citep{Zucca1997,Blanton2005b} and in clusters and groups
\citep{Trentham1998,Trentham2005,Popesso2007,Jenkins2007}.  The
departure of the GSMF from a single Schechter function at low stellar
masses was noticed by \citet{Baldry2006,Baldry2008} and
\citet{Panter2004} for SDSS data. At higher redshifts, an \emph{a
posteriori} look at the published GSMFs often reveals such an upturn.
 
We refer to $\mathcal{M}_{\rm min}$ as the lowest mass at which the
GSMF can be considered reliable and unaffected by incompleteness on
$\mathcal{M}/L$ \citep[see][]{Ilbert2004,Pozzetti2007}. A complete
description of the procedure can be found in \citet{Pozzetti2010}.
Our aim is to recover the stellar mass up to which all the galaxy
types contributing significantly to the GSMF can be observed. We
derived this value in small redshift slices by considering the $20$\%
faintest galaxies, i.e., those contributing to the low-mass end of the
GSMF.  For each galaxy of this subsample, we computed the ``limiting
mass'', that is the stellar mass that the object would have had at the
limiting magnitude of the survey, $\log\mathcal{M}_{\rm lim} =
\log\mathcal{M} + 0.4(I-22.5)$.  For each redshift bin, we define as
minimum mass the value corresponding to $95$\% of the distribution of
limiting masses and we smooth the $\mathcal{M}_{\rm min}$ versus $z$
relation by means of an interpolation with a parabolic curve.  The
minimum stellar mass we adopt is the value up to which we can reliably
compute the GSMF in each considered redshift bin, i.e. the
$\mathcal{M}_{\rm min}$ at the lowest extreme of the interval, since
the $1/V_{\rm max}$ method corrects the residual volume
incompleteness.

We note that this limit substantially decreases the number of objects
considered in each redshift bin to derive the GSMF.  The redshift
intervals $[0.10,0.35]$, $[0.35,0.50]$, $[0.50,0.70]$, and
$[0.70,1.00]$ were chosen to contain a similar number of galaxies and
the values we obtained for the limiting mass of the total sample are
$\log\mathcal{M}_{\rm lim}/\mathcal{M}_\odot = 8.2, 9.4, 9.9, 10.5$
from the lowest to the highest redshift bin.  When dealing with GSMFs
divided into galaxy types, the minimum masses are obtained separately
for each subsample.

\subsection{The choice of the environment definition}
\label{envchoice}

As mentioned in Sect.~\ref{env}, the density field of the COSMOS field
\citep[see][]{Kovac2010a} was reconstructed for different choices
of filters (of fixed comoving aperture or adaptive with a fixed number
of neighbours), tracers (from flux-limited or volume-limited
subsamples), and weights (stellar mass, luminosity or no weight, i.e.,
considering only the number of galaxies).

We tested the options that allow an unbiased comparison over the whole
redshift range, from $z=0.1$ to $1.0$.  In particular, we explored the
5NN estimator and the 5NN mass-weighted one (hereafter 5NNM), both of
them computed using volume-limited tracers, with two choices of
luminosity limits: $M_B \le -20.5-z$ (bright tracers) and $M_B \le
-19.3-z$ (faint tracers), where $M_B$ is the absolute magnitude in the
$B$ band computed with ZEBRA.  The absolute magnitude cut was derived
by considering the distribution of absolute magnitudes versus
redshift, the so-called Spaenhauer diagram \citep{Spaenhauer1978}, and
the evolution of the parameter $M_B^*$ of the LFs \citep{Zucca2009}.
Two different limits are necessary because of the rareness of bright
tracers at low redshift and the incompleteness of faint tracers at
high redshift; for this reason, the two overdensity estimates cannot be
computed over the whole redshift range, but only at $[0.1,0.7]$ and
$z=[0.4,1.0]$ for faint and bright tracers, respectively.

\subsubsection{The effect of environment tracers on GSMF}

Two problems affect the study on the evolution of GSMFs as a function
of environment, which must be solved: (1)~we have to understand
whether the 5NNM estimator is a more robust tracer of the environment,
as predicted theoretically; (2)~we have to be certain that the use of
two different tracers, e.g., with a change at $z=0.7$, does not
introduce a spurious signal that may be misinterpreted as an
evolutionary trend.

To answer both questions, we used as a test case the redshift interval
$[0.4,0.7]$, where all the estimates are available, and we computed
the quartiles of the $1+\delta$ distribution in this redshift bin
considering only the objects with masses higher than the minimum
mass. Henceforth, we refer to the lowest and highest quartiles of
$1+\delta$ as D1 and D4, respectively. In the reminder of the paper,
we focus our study on these two extremes.

\begin{figure}
\centering
\includegraphics[width=\hsize]{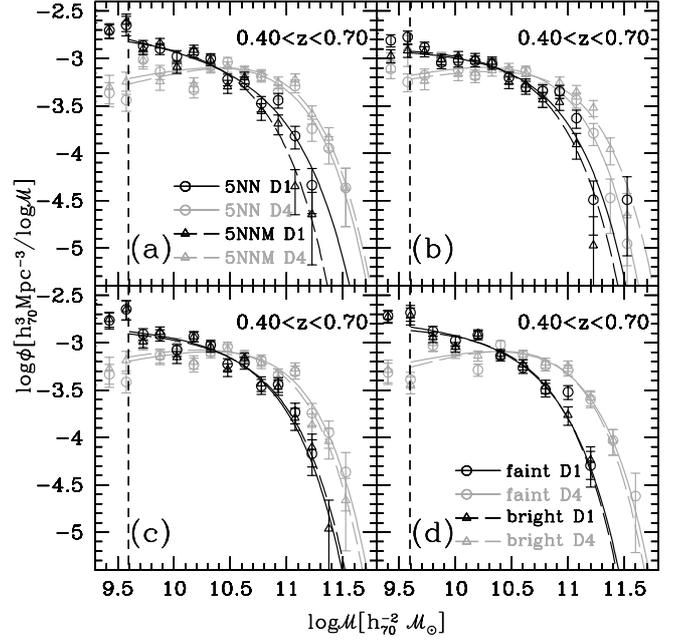}
\caption{
(a) Comparison of GSMFs for environment estimates from 5NN and 5NNM
volume limited with faint tracers: Black: D1 (underdense); Grey: D4
(overdense). Solid line and empty dots: 5NN. Dashed line and empty
triangles: 5NNM. The vertical dashed line represent the value of
$\mathcal{M}_{\rm min}$ at $z=0.4$.
(b) As in panel (a), but 5NN and 5NNM overdensities have been
estimated after a random shuffling of galaxy properties to remove the
mass--density relation.
(c) As in panel (a), but 5NN and 5NNM overdensities have been
estimated without considering the properties of the central galaxy.
(d) GSMFs for bright ($M_B \le -20.5-z$, dashed lines and empty
triangles) and faint ($M_B \le -19.3-z$, solid lines and empty dots)
tracers using 5NN overdensities in the D1 (black) and D4 (grey)
environments.}
\label{fig:mf_test_5NN}
\end{figure}

In Fig.~\ref{fig:mf_test_5NN}, panel (a), we compare the GSMFs derived
using a single Schechter function fit, for 5NN and 5NNM overdensity
estimators, both using the faint volume-limited tracers. The
separation of GSMFs between D1 and D4 environments is more prominent
when considering the mass-weighted estimator, because of the larger
dynamical range of the $1+\delta$ values studied. In particular, the
main difference is in the massive part of D1 GSMF: massive galaxies in
low density environments using 5NN move to intermediate densities for
5NNM estimator because of their high stellar masses.  This decreases
the number (and therefore the normalisation of the GSMF) of massive
galaxies in low density environments when the 5NNM estimator is
adopted. To test whether this enhancement of the difference between
the D1 and D4 GSMFs is real, we performed the following test: we
removed as much as we could the mass--density relation by shuffling
the original catalogue and computing overdensities considering objects
with their original coordinates, but assigning to each one the
observed properties (magnitudes, stellar mass, weight) of the 25th
following object after redshift sorting. Both 5NN and 5NNM
overdensities and their quartiles were then recomputed, since the
shuffling also changes the tracers.  The choice of the 25 object jump
is a compromise between the requirements of preserving a similar
probability of being observed at the chosen redshift (i.e., avoiding
unphysical galaxy properties if a large jump in redshift is allowed)
and selecting objects possibly not in the same structure, where we
know galaxies share similar properties (in this case the mass--density
relation would not be removed).  In this test, we expect that GSMFs
derived in D1 and D4, regardless of the estimator of the density
contrast used, to be approximately the same, since, after reshuffling,
massive galaxies should no longer occupy preferentially high density
environments.  Moreover, we also expect that the 5NN and 5NNM
estimators of the density should produce similar results, since the 5
neighbours should have a random distribution of their stellar masses.
The comparison between GSMFs with 5NN and 5NNM ``shuffled''
overdensities is shown in Fig.~\ref{fig:mf_test_5NN}, panel (b).  For
5NN, we see that GSMFs in D1 and D4 are more similar than before, but
not coincident; this may be due to an insufficiently large amount of
shuffling being used to separate masses and environment in the biggest
structures.  Furthermore, the 5NN and 5NNM estimates are still quite
different, mainly at the high masses in D4. These results may be
caused by the non-negligible influence of the stellar mass of the
object itself in the case of the 5NNM estimator, possibly enhanced by
a residual signal in the mass--density relation.

In our last test to interpret this residual signal, we removed the
central galaxy when computing $1+\delta$ from the original catalogue:
the comparison of the resulting GSMFs is in Fig.~\ref{fig:mf_test_5NN}
panel (c), which shows now fully consistent GSMFs at high and low
densities as defined from 5NN and 5NNM estimators.

These tests seem to indicate that the mass weighting scheme assigns
too great an importance to stellar masses on scales of the order of
the galaxy itself.  Thus, we attempted to avoid any possible bias due
to stellar mass over-weighting, despite its physically motivated link
with the halo mass, by discarding the 5NNM estimator and we performed
our analysis using number-weighted overdensities.

To help resolve the second problem, we tested whether the change of
the tracers at $z=0.7$ could introduce some change in the GSMF, which
can be misinterpreted as evolution. We already know that the scales
probed at the same $1+\delta$ are more or less twice as large for
bright than faint tracers \citep{Kovac2010a}, therefore it is not
possible to use the same $1+\delta$ threshold for both faint and
bright tracers. To overcome this problem, we determined the quartiles
of $1+\delta$ separately for each redshift bin. The results of this
test are shown in panel (d) of Fig.~\ref{fig:mf_test_5NN}. In the
$z=0.4-0.7$ bin, where both tracers are available, the GSMFs
obtained with the two tracers, with independently computed quartiles,
are completely consistent with each other in under and overdense
environments D1 and D4, and therefore we assume we can safely compare
the results at redshifts $z<0.7$ computed with faint tracers to those
computed at $z\ge 0.7$ with the bright ones.

\subsubsection{Definition of overdensity quartiles}
\label{defD1D4}

\begin{figure}
\centering
\includegraphics[width=\hsize]{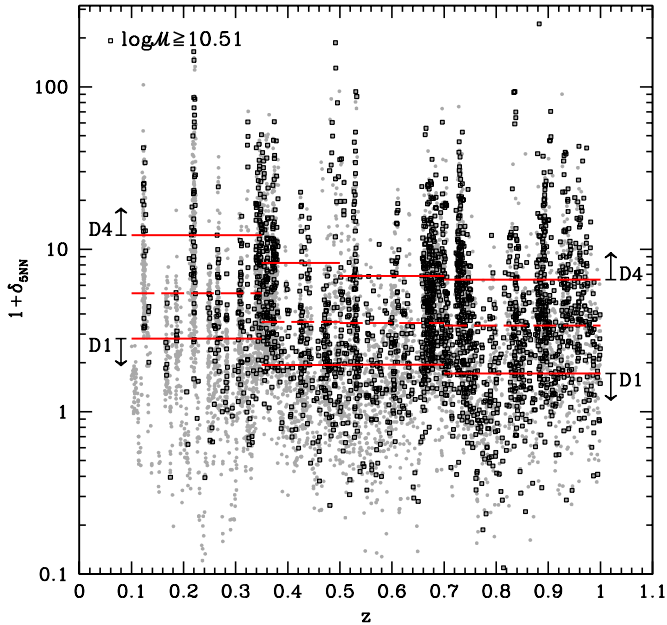}
\caption{
Definition of quartiles for the 5NN estimator using volume-limited
tracers: grey points represent the full sample, black squares the
galaxies with masses above the $\mathcal{M}_{\rm min}$ computed in the
last redshift bin, horizontal segments show the values of the
quartiles of $1+\delta$ computed from the distribution of these
massive galaxies, and the dashed ones indicate the median.}
\label{fig:quartiles}
\end{figure}

As already mentioned, we traced the effect of extreme environments on
the evolution of galaxies by considering the quartiles D1 and D4 of
the $1+\delta$ distribution, using 5NN volume-limited overdensities.
The quartiles were computed at each redshift bin considering only
the population of galaxies more massive than the minimum stellar mass
considered for the GSMF (see Sect.~\ref{mf_method}) in the highest
redshift bin, i.e., ${\log \mathcal M}_{\rm min}/\mathcal{M}_\odot
\simeq 10.5$, to ensure that this definition is unaffected by the
variation as a function of redshift in the observable mass range,
populated by different mix of galaxy types.  The quartile definition
used throughout this paper is shown in Fig.~\ref{fig:quartiles}. The
median scales probed by the $5$th nearest neighbour range from
$0.87\,{\rm Mpc}\,h_{70}^{-1}$ in the D4 environment at low redshift
to $7.57\,{\rm Mpc}\,h_{70}^{-1}$ in the D1 quartile at the highest
redshift bin, where we have to use bright tracers.

The trend toward higher values of overdensity at lower redshifts is in
some measure expected from the growth of structures, which amplifies
the dynamic range of overdensities, but this increase cannot be
quantified using the linear approximation, which is invalid on the
scales probed by our density estimates.  The different values of the
$1+\delta$ quartiles in the different redshift bins correspond to very
similar scales when the same tracers are used.

It is not easy to compare the values of density contrast in
Fig.~\ref{fig:quartiles} with those of known objects, such as rich
clusters or voids, because of the different definitions of environment
and the different scales probed. A possible comparison is instead
feasible with the distribution of $1+\delta$ for the members of galaxy
groups identified in the same COSMOS field. This comparison is shown
in Fig.~22 of \citet{Kovac2010a}, where it is possible to see that
galaxy members of optical groups with $\ge 2$ members have a
distribution of overdensities that peaks at $1+\delta \sim 6$, whereas
richer groups and X-ray candidate clusters typically have $1+\delta
\sim 20$.  Although the different classifications of the environment
are obviously related, they are not perfectly coincident, with $\sim
59$\,\% of the objects in the group catalogue used by
\citet{Kovac2010b} belonging to D4 (and only $6$\% to D1) and $\sim
73$\,\% of the objects classified as ``isolated'' by
\citet{Iovino2010} being in D1 (and only $0.2$\% in D4).

\subsection{Mass functions in different environments}
\label{mfenv}

\begin{figure}
\centering
\includegraphics[width=\hsize]{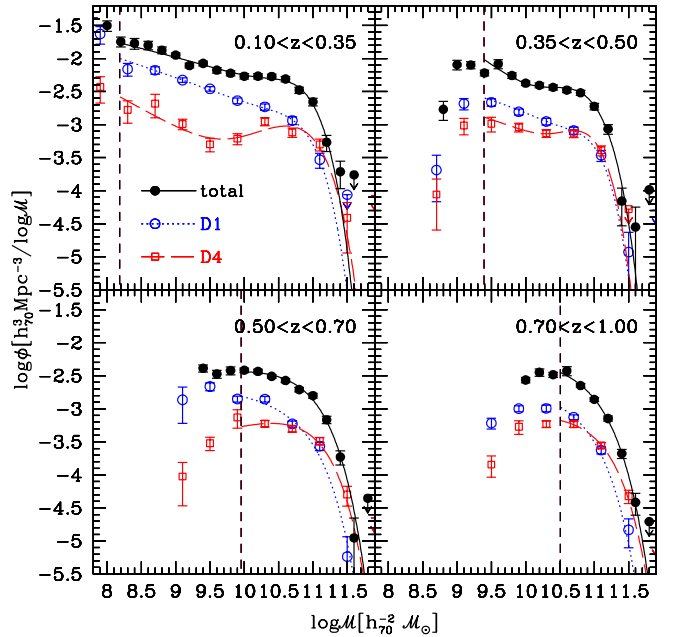}
\caption{
The MFs in the extreme quartiles D1 and D4 of the 5NN volume-limited
overdensities. Black: total GSMF, with $1/V_{\rm max}$ dots and their
Poissonian error bars and Schechter function fit (double Schechter
function in the first two redshift ranges and a single one at higher
redshifts). Blue: lowest $1+\delta$ quartile. Red: highest density
quartile. }
\label{fig:mfenv}
\end{figure}

The GSMFs in the two extreme environments are shown in
Fig.~\ref{fig:mfenv}: the bimodality is clearly visible in the global
GSMFs \citep[][see also the points and lines in
Fig.~\ref{fig:mfenv}]{Pozzetti2010}, with an upturn at the low-mass
end around $\mathcal{M}\sim 10^{9.5}\,\mathcal{M}_{\odot}$, which is
more pronounced in the high density regions, at least in the two
lowest redshift bins.  We used the double Schechter function fit only
up to $z\sim 0.5$, where the dip in the GSMFs falls at stellar masses
higher than $\mathcal{M}_{\rm min}$.  Because of our choice of
environment definition, the normalisation of D1 and D4 GSMFs does not
have a clear physical meaning, since the volumes occupied by each
galaxy are referred to the total volume of the survey and the number
of galaxies in each environment is not $1/4$ of the total sample.  To
obtain a more meaningful definition of the normalisation, we should
compute the volume occupied by the structures with the considered
ranges of $1+\delta$; here we compare only the GSMF shapes, hence
defer a more in-depth study of the normalisation to a future
analysis. A striking difference in GSMF shapes is evident, with
massive galaxies preferentially residing in high density environments,
characterised on average by a higher $\mathcal{M}^*$, and with a
steeper slope than the D1 GSMFs at $z\ge 0.35$.  The different shapes
and the strong bimodality in the D4 GSMF can be interpreted in a
similar way to the global one \citep{Pozzetti2010} by the different
contribution of different galaxy types, as we see in the next section.
The parameters of the Schechter fits to the GSMFs are listed in
Table~\ref{tab:mfenv}.

\begin{table}
\caption{Parameters of the GSMF in the low and high-density environments.}
\begin{tabular}{lccccc}
\hline\hline
   & $z$ & $\alpha_1$ & $\alpha_2$ & $\log\mathcal{M}^*/\mathcal{M}_\odot$ &
   $\phi^*_1/\phi^*_2$\\
\hline
D1 & $0.10-0.35$ &  -1.35 &  +0.14 & 10.53  & 1.61 \\
   & $0.35-0.50$ &  -1.25 &  +0.82 & 10.52  & 0.79 \\
   & $0.50-0.70$ &  -1.13 &   ...  & 10.82  & ...  \\
   & $0.70-1.00$ &  -1.12 &   ...  & 10.80  & ...  \\
\hline
D4 & $0.10-0.35$ &  -1.80 &  -0.33 & 10.76  & 0.01 \\
   & $0.35-0.50$ &  -1.28 &  +0.95 & 10.52  & 0.50 \\
   & $0.50-0.70$ &  -0.70 &   ...  & 10.92  & ...  \\
   & $0.70-1.00$ &  -0.90 &   ...  & 10.98  & ...  \\
\hline
\end{tabular}
\label{tab:mfenv}
\end{table}


\section{Evolution of galaxy types in different environments}
\label{mftype}

\begin{figure*}
\centering
\includegraphics[width=0.48\hsize]{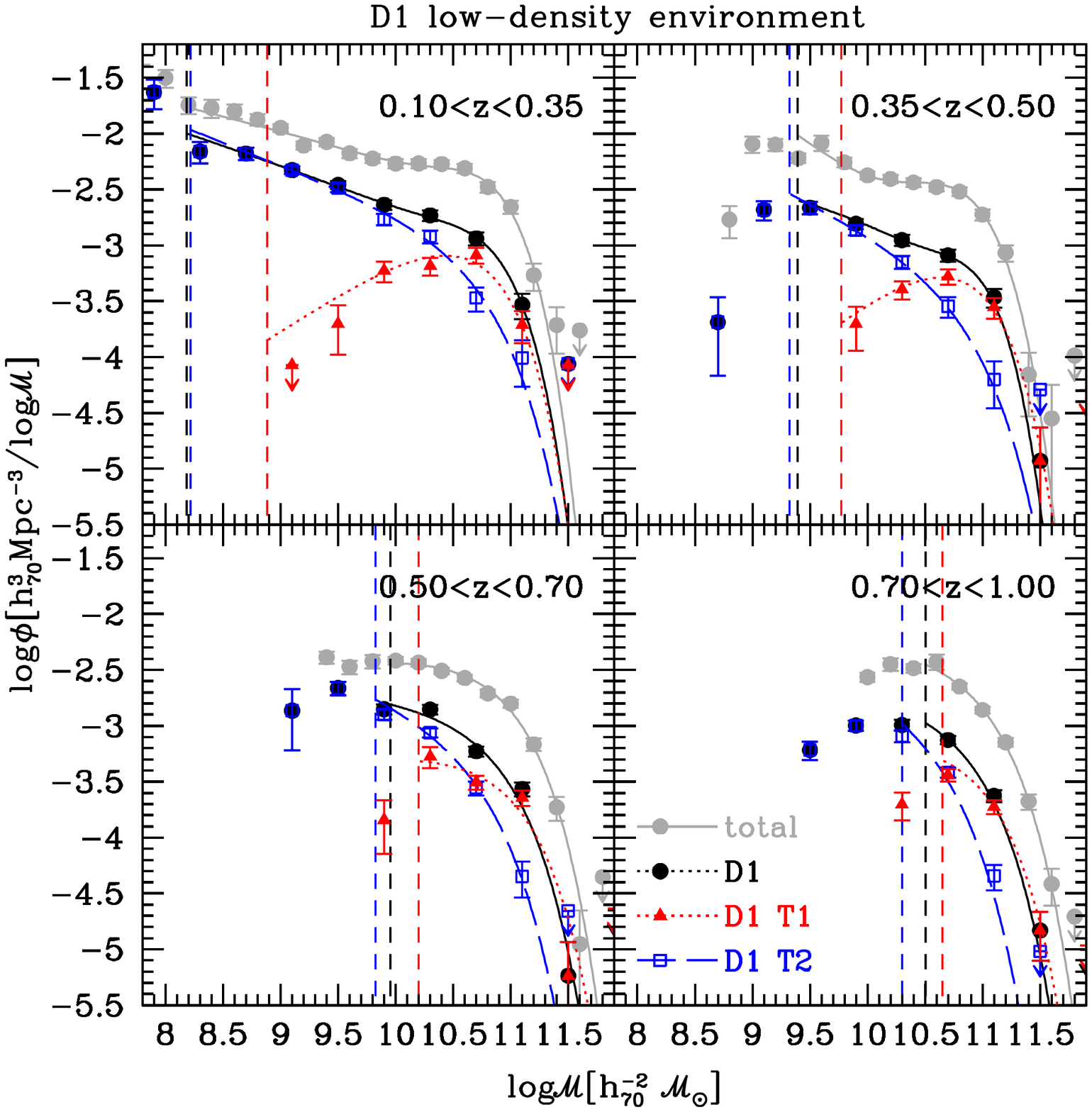}
\includegraphics[width=0.48\hsize]{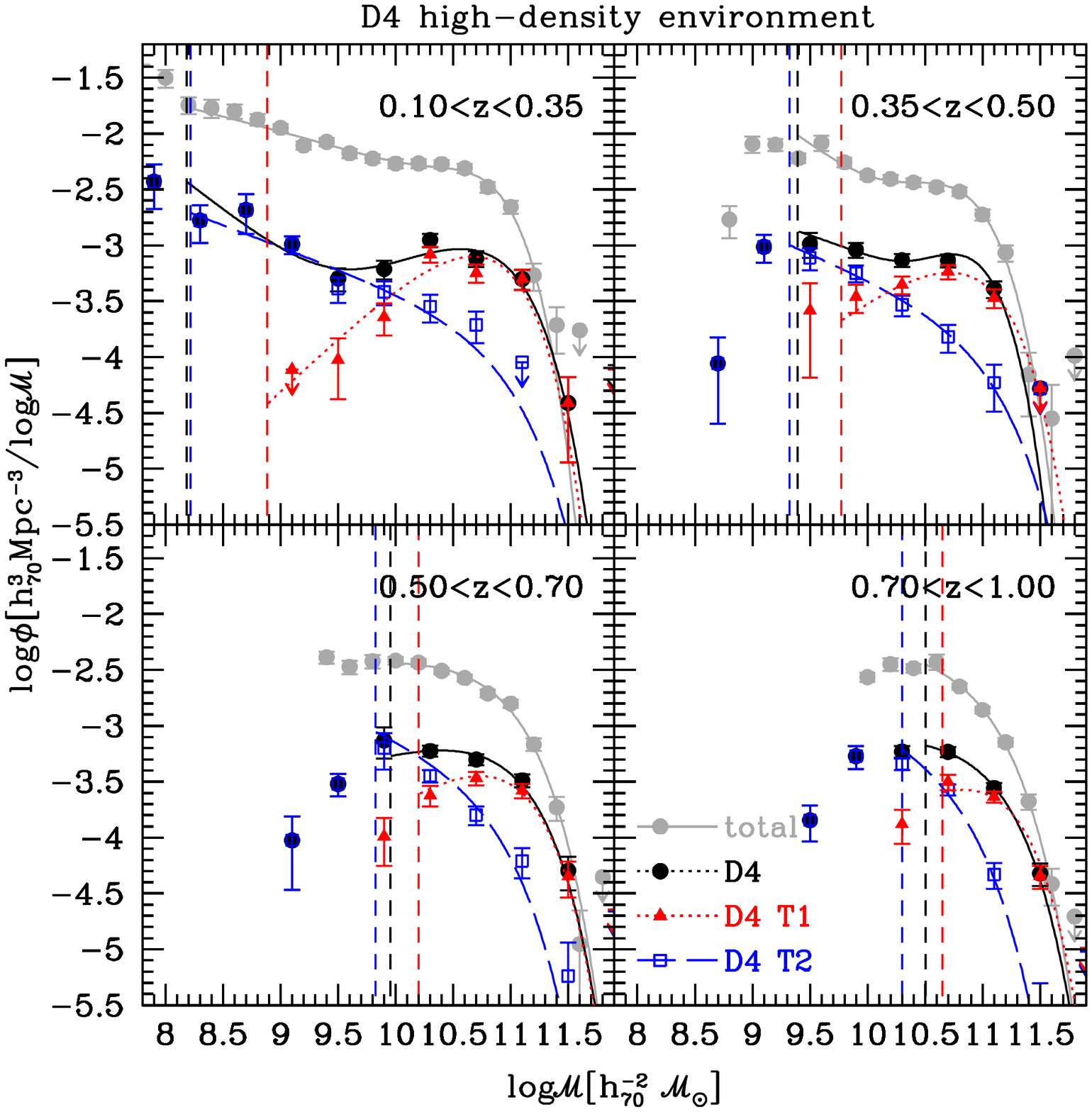}
\caption{
Left: quartile D1 (low density environment). Right: quartile D4 (high
density).  Grey: total GSMF. Black: MF relative to the considered
quartile. Red triangles and dotted lines: photometric early-type
galaxies. Blue squares and dashed lines: photometric late-type
galaxies. At high masses, the upper limit points show the $2\sigma$
confidence limits for $0$ detections following \citet{Gehrels1986}.}
\label{fig:mf_delta_type}
\end{figure*}

The need to use a double Schechter function to fit the global and
environment-selected GSMFs at least up to $z\sim 0.5$ may be linked to
the contribution of different galaxy populations. Galaxies with the
same luminosity may be characterised by very different
$\mathcal{M}/L$, which can explain why it is difficult to identify the
bimodal shape of LFs, even though this bimodality was first detected
in LFs.

To study the contribution of galaxies with different photometric types
and morphologies in the extreme environments, we computed the GSMFs of
D1 and D4, defined as in Sect.~\ref{mfenv}, by dividing each
sub-sample into galaxy classes.  The values of $\mathcal{M}_{\rm min}$
were computed separately for early/elliptical/bulge-dominated and
late/spiral/disc-dominated galaxies.  These values differ
significantly, especially at low redshift, confirming the very
different distributions of $\mathcal{M}/L$.

The results for the contribution of different photometric types to D1
and D4 GSMFs are presented in Fig.~\ref{fig:mf_delta_type}, and the
best-fit parameters of the single Schechter function fits are given in
Table~\ref{tab:mfenvt}. Dividing the sample into the two broad
morphological classes results in qualitatively similar GSMFs.

Looking at the plots in Fig.~\ref{fig:mf_delta_type}, it is clear that
the stronger bimodality in the first two redshift bins in the D4 GSMF
is primarily due to the larger contribution of early-type galaxies. As
for the global GSMF, in both of the considered environments early-type
galaxies are dominant at high masses ($\log\mathcal{M/M}_\odot \ga
10.7$), while their contribution rapidly decreases at intermediate
masses. On the other hand, late-type galaxies, which have much steeper
GSMFs, start to dominate at intermediate and low masses
($\log\mathcal{M/M}_\odot \sim 10$).

\begin{figure*}
\centering
\includegraphics[width=0.48\hsize]{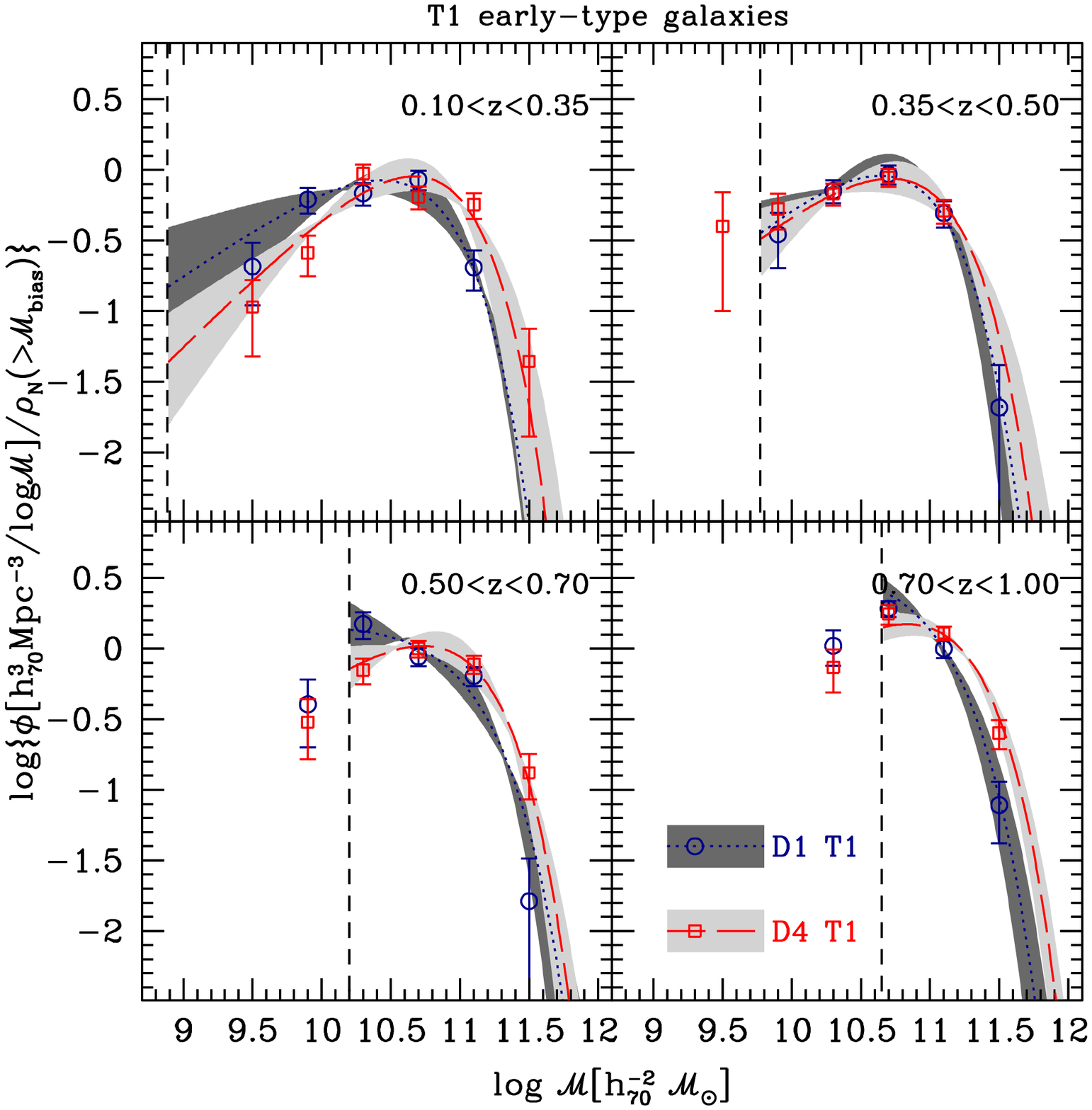}
\includegraphics[width=0.48\hsize]{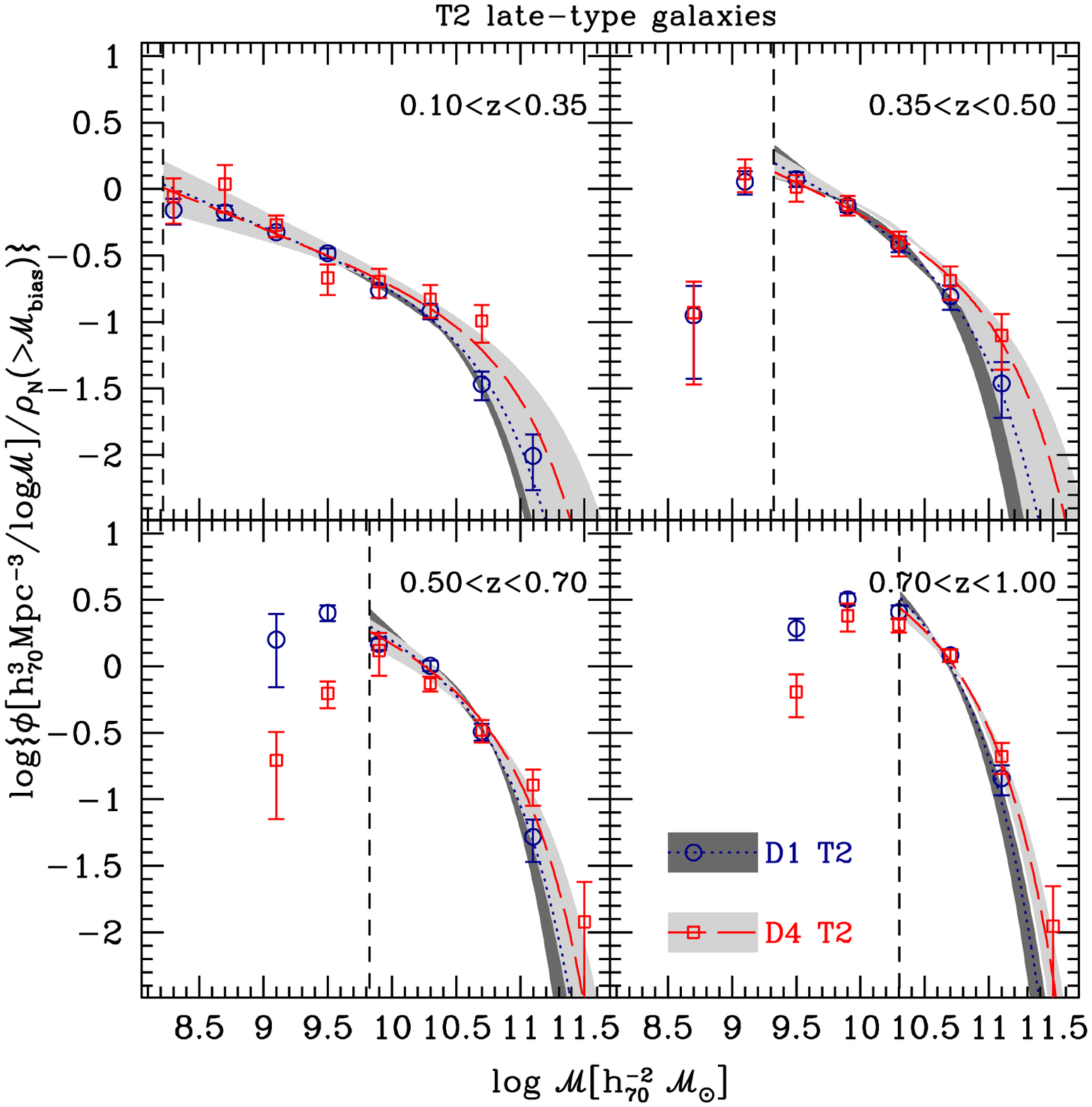}
\caption{
Left: GSMFs of photometric early-type galaxies in D1 and D4
environments, renormalised to number density $=1$ for stellar masses
$>\mathcal{M}_{\rm min}$. Right: the same for photometrically late
type galaxies. Dotted lines, circles and dark shaded regions represent
the GSMFs in underdense regions, D1. Dashed lines, squares and light
shaded regions illustrate D4 GSMFs.}
\label{fig:mf_type_d14}
\end{figure*}

In addition to assessing the relative contributions of different
galaxy types in D1 and D4, it is sensible to ask whether the shape of
the GSMFs of galaxies of the same type is the same in different
environments, i.e., whether a ``universal'' mass function of
early/late-type galaxies does exist.  In Fig.~\ref{fig:mf_type_d14},
we compare early- and late-type GSMFs in the two environments, in each
redshift bin renormalised with the number density computed for masses
$\ge \mathcal{M}_{\rm min}$.  The shapes of the GSMFs differ slightly,
there being a slightly higher density of massive galaxies in overdense
regions; however the similarity of the GSMFs in all the redshift bins,
and in particular for late-type galaxies, is remarkable and somewhat
unexpected.  If the shape of the GSMF of galaxies of the same type is
similar in different environments, any difference seen in the total
GSMFs in under- and overdense regions at low redshift should be due to
the different evolution of their normalisations.

\begin{figure}
\centering
\includegraphics[width=\hsize]{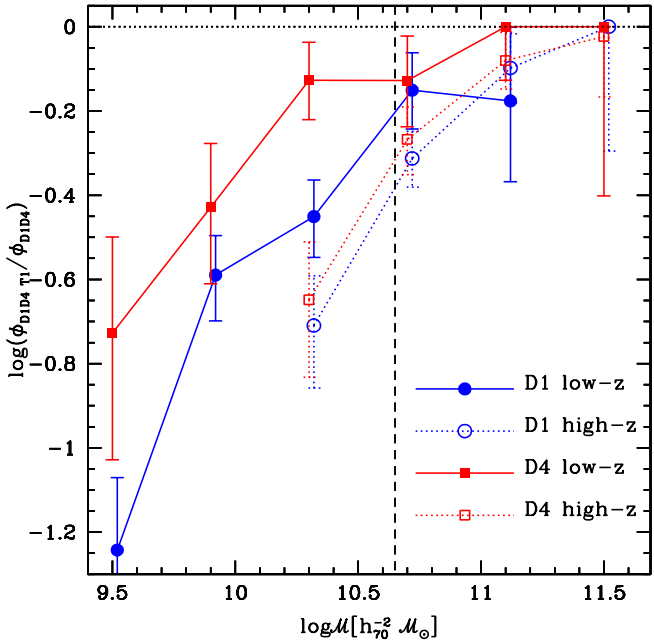}
\caption{
Evolution of the fractional contribution of the photometric early-type
to the global MFs (the late-type fractional contribution is
complementary to the one shown in this plot) in the two extreme
environments. Blue lines and circles refer to the low density
environment D1 (displaced by 0.02 in the abscissa to avoid
overlapping), red lines and squares to the high density sample
D4. Dotted lines and empty symbols represent the highest redshift bin
$z=[0.7,1.0]$, solid lines and filled points the lowest one,
$z=[0.1,0.35]$.  The vertical dashed line indicates $\mathcal{M}_{\rm
min}$ in the high redshift bin (the value at low redshift is outside
the plot).  Error bars have been computed as $16-84$\% of the
distribution of Monte Carlo simulations.}
\label{fig:frac_mf_delta_type}
\end{figure}

\begin{table}
\caption{Parameters of the GSMF for the two photometric types T1 
(early-type galaxies) and T2 (late-type galaxies) in the low and
high-density environments. When the parameter $\alpha$ is
undetermined, we fixed it to the best-fit value in the previous bin of
the same environment. Error bars are at $1\sigma$ confidence level.}
\begin{tabular}{lccc}
\hline\hline
   & $z$ & $\alpha$ & $\log\mathcal{M}^*/\mathcal{M}_\odot$ \\
\hline
D1T1 & $0.10-0.35$ & $-0.33^{+0.46}_{-0.37}$  & $10.60^{+0.15}_{-0.11}$\\
     & $0.35-0.50$ & $-0.17^{+0.71}_{-0.55}$  & $10.72^{+0.32}_{-0.21}$\\
     & $0.50-0.70$ & $-0.90^{+0.85}_{-0.60}$  & $10.93^{+0.26}_{-0.25}$\\
     & $0.70-1.00$ & $[-0.90]$                & $10.88^{+0.10}_{-0.10}$\\
\hline
D1T2 & $0.10-0.35$ & $-1.41^{+0.11}_{-0.07} $ & $10.71^{+0.18}_{-0.23}$\\ 
     & $0.35-0.50$ & $-1.51^{+0.32}_{-0.25} $ & $10.81^{+0.51}_{-0.36}$\\
     & $0.50-0.70$ & $-1.45^{+0.52}_{-0.36} $ & $10.70^{+0.28}_{-0.26}$\\
     & $0.70-1.00$ & $[-1.45]$                & $10.59^{+0.06}_{-0.08}$\\
\hline
D4T1 & $0.10-0.35$ & $-0.03^{+0.46}_{-0.32}$  & $10.68^{+0.18}_{-0.21}$\\
     & $0.35-0.50$ & $-0.23^{+0.59}_{-0.45}$  & $10.82^{+0.22}_{-0.20}$\\
     & $0.50-0.70$ & $-0.28^{+0.65}_{-0.48}$  & $10.87^{+0.15}_{-0.18}$\\
     & $0.70-1.00$ & $[-0.28]$                & $10.97^{+0.06}_{-0.06}$\\
\hline
D4T2 & $0.10-0.35$ & $-1.39^{+0.13}_{-0.09}$  & $10.92^{+0.32}_{-0.38}$\\
     & $0.35-0.50$ & $-1.43^{+0.19}_{-0.14}$  & $11.02^{+0.22}_{-0.45}$\\
     & $0.50-0.70$ & $[-1.43]$                & $10.81^{+0.13}_{-0.15}$\\
     & $0.70-1.00$ & $[-1.43]$                & $10.75^{+0.07}_{-0.07}$\\
\hline
\end{tabular}
\label{tab:mfenvt}
\end{table}

To examine the differential contribution of various galaxy types in
different environments, we can compute the evolution of the ratio of
the GSMF of a given galaxy class to the global GSMFs in each
environment. In Fig.~\ref{fig:frac_mf_delta_type}, we show the ratio
of $1/V_{\rm max}$ estimates of early-type GSMF in over and underdense
regions for the two extreme redshift bins. The trend for late-type
galaxies is the opposite of that shown in the figure.  The error bars
were computed using a Monte Carlo simulation considering Gaussian
distribution of errors of rms derived from Poissonian error bars using
$1/V_{\rm max}$ method. The $16$\% and $84$\% of the $100\,000$
iterations of the ratio distribution are reported in the plot as error
bars.  The vertical dashed line shows the value of $\mathcal{M}_{\rm
min}$ for early-type galaxies in the redshift bin $z=0.7-1.0$. Despite
the large error bars, Fig.~\ref{fig:frac_mf_delta_type} illustrates
that in the high redshift bin the fractional contributions of
photometric early-types to the GSMF in different environments are more
or less the same for D1 and D4 at all the masses we can safely study.
On the other hand, the fractional contribution is significantly
different at low redshift, mainly at intermediate stellar masses
($\log\mathcal{M/M}_\odot \la 10.5$).  This trend appears to imply
that there is a more rapid growth with time in high density
environments of the fractional contribution of early-type galaxies.
At intermediate masses, the differences between the two extreme
environments are larger: high stellar masses ($\log\mathcal{M/M}_\odot
\ga 10.7$) are populated mainly by passive red galaxies in both
environments, while at lower masses ($\log\mathcal{M/M}_\odot \la 10$,
in the low redshift bins, where it is possible to probe them) the
population of late-type/star-forming galaxies dominates in all the
environments.

\begin{figure}
\centering
\includegraphics[width=\hsize]{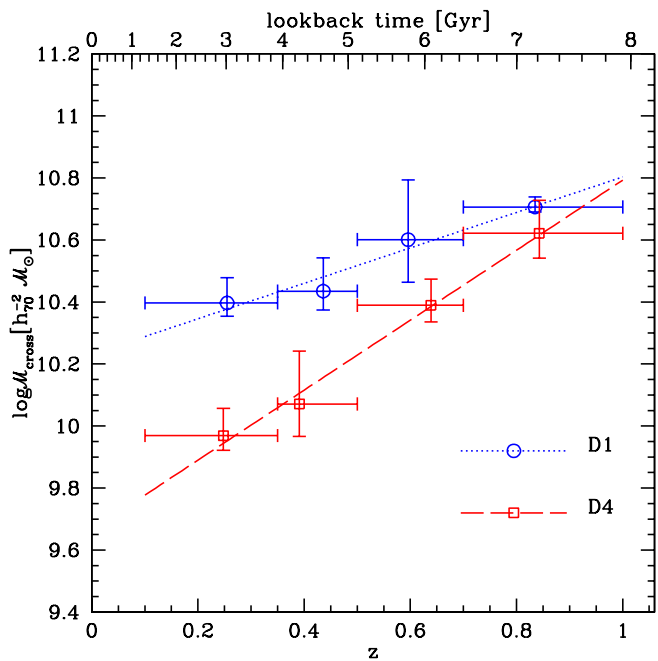}
\caption{
${\mathcal M}_{\rm cross}$ of photometric types in the extreme
quartiles D1 and D4. Blue: low-density environments. Red:
high-density. The points are located at the median redshift of the
early plus late samples and error bars represent the width of the
redshift bin and the error in the GSMF ratio from $1/V_{\rm max}$
method. A linear fit to the points is also shown.}
\label{fig:mcrosst}
\end{figure}

In a scenario that is consistent with these data, which indicate there
is an increase in early-type galaxies with cosmic time,
blue intermediate-mass galaxies are being transformed into more
massive red galaxies, after quenching their star formation in a more
efficient way in overdense than underdense regions.  A possible way to
quantify this difference in evolutionary speed is by analysing the
evolution with redshift of ${\mathcal M}_{\rm cross}$, which
represents the mass above which the GSMF is dominated by early-type
galaxies.  We show this quantity computed from $1/V_{\rm max}$ points
in Fig.~\ref{fig:mcrosst} for different photometric types.  We can see
that since $z\sim 1$, where the $\mathcal{M}_{\rm cross}$ values in
low and high density environments were similar, the subsequent
evolution produces a significant difference between the two
$\mathcal{M}_{\rm cross}$ values.  The ratio of $\mathcal{M}_{\rm
cross}$ in the highest to lowest redshift bins implies an evolution of
a factor $\sim 2$ in low density and $\sim 4.5$ in high density
regions.  From a different point of view, the plot in
Fig.~\ref{fig:mcrosst} indicates that the environment begins to affect
the evolution of galaxies at $z\sim 1$, causing in the lowest redshift
bin a delay of $\sim 2$\,Gyr in underdense relative to overdense
regions before the same mix of galaxy types is observed in high
density regions.

\section{Discussion}
\label{discuss}

\subsection{Comparison with literature data}
\label{literature}

As mentioned in Sect.~\ref{intro}, a similar analysis of the influence
of environment on the evolution of the GSMF of red and blue galaxies
was carried out by \citet{Bundy2006} using DEEP2 data.  They
considered a sample in the redshift range $0.4<z<1.4$, partially
overlapping with ours, and a the definition of galaxy types and
environment that slightly differed; their galaxy types are defined on
the basis of the rest-frame colour $U-B$ and their under- and
overdense environments are defined with respect to the average local
density for the majority of their analysis.  Since, as the authors
also state, most of the galaxies belong to regions around the average
density, we do not expect to find that the environment has a
significant influence of the redshift evolution of galaxies.  However,
they also considered the extremes of the density field in their
Fig.~11, where they present the evolution with redshift in the
fractional contribution of red and blue galaxies.

\begin{figure}
\centering
\includegraphics[width=\hsize]{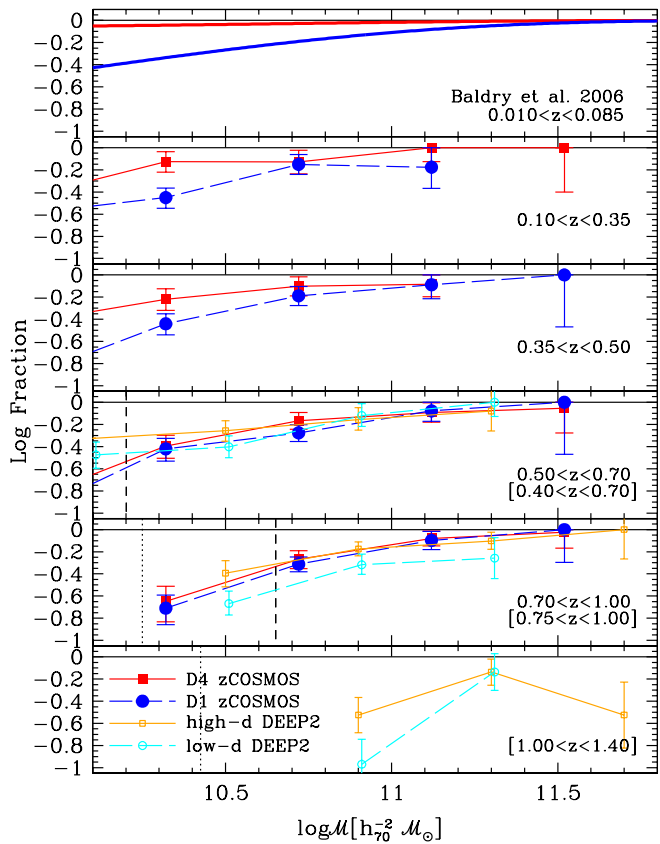}
\caption{
Evolution of the fractional contribution of the early-type/red
galaxies to the global MFs in low and high density environments from
the surveys SDSS, zCOSMOS, and DEEP2.  In the low redshift bin, red and
blue lines are computed from Eq.~10 by \citet{Baldry2006},
representing the fraction of red galaxies in the highest and lowest
environmental densities in their SDSS analysis.  In the other
redshift bins, red solid lines and filled squares represent the
zCOSMOS high-density sample D4, and blue long-dashed lines and filled
circles the low-density sample D1. Orange and cyan lines and empty
symbols represent the values of the analogous fractions taken from
\citet{Bundy2006}.  The vertical dashed lines mark $\mathcal{M}_{\rm
min}$ in zCOSMOS, and vertical dotted lines represent the $K_s$-band
completeness limits in \citet{Bundy2006}. Redshift ranges between
brackets refer to DEEP2 binning.  }
\label{fig:frac_bundy}
\end{figure}

We compare our results obtained using our definitions of environment
and galaxy types, with the \citet{Bundy2006} paper in
Fig.~\ref{fig:frac_bundy}.  At low redshift, we plot for reference the
results of \citet{Baldry2006}, who used SDSS data divided into density
bins and galaxy types separated by means of the colour bimodality. The
lines in the plot are derived from their Eq.~10, adopting their
highest and lowest density values.  The results from the two high-$z$
surveys are in reasonably good agreement. The largest difference is in
low density environments in our redshift bin $z=[0.70,1.00]$, but
results are marginally consistent with each other.  When we study the
evolution of the mass function fractions derived from the three
surveys, the main visible trend is the continuous increase with time
in the fractional contribution of red/early-type galaxies in all
environments, which is an alternative way of observing the build-up of
the red sequence and its increasing population at lower stellar
masses.  The differences between low and high density environments
seem to increase towards low redshift, whereas at high redshifts the
quite large error bars prevent our drawing robust conclusions, which
may also depend on the particular definitions of the samples.

\citet{Cooper2010} analysed the colour--density relation in the DEEP2 
sample and claimed that the environmental dependence is still present
at $z=1$.  In contrast to our analysis, they considered the top $10$\%
of the high-density sample, using the density field computed at the
distance of the 3rd-nearest neighbour in the total flux-limited
sample.  With this choice, they explored a smaller scale environment
than the one used in the present paper. For instance, they state that
the typical distance involved in the computation of their top $5$\%
overdensities is about $35\arcsec$ at $z\sim 0.9$, corresponding to a
comoving scale $\sim 0.37\,h^{-1}$\,Mpc. The average scale of our top
$5$\% overdensities in our highest redshift bin is $\sim
1.1\,h^{-1}$\,Mpc.  Therefore, the results of the two surveys do not
necessarily disagree if the environmental mechanism modifying galaxy
properties at $z \ga 1$ is mainly effective on small scales.

Other studies of the evolution of the GSMFs of galaxies of different
types and morphologies are presented in \citet{Pozzetti2010},
\citet{Ilbert2010}, \citet{Bundy2010}, and \citet{Drory2009}, though
without incorporating directly the environmental effects. They all
find that the global GSMF has a bimodal shape, with the need to use
two Schechter functions eventually extending to the single galaxy
types GSMFs, as found by \citet{Drory2009}. These authors interpret
the presence of a plateau at $\sim 10^{10} \mathcal{M}_\odot$ in blue
galaxies as a signature of either a change in star formation
efficiency, which is more dramatic at lower masses, or an increase in
the galaxy assembly rate at higher masses.  At low redshift, the dip
appears to move from blue to red galaxies, because blue massive
galaxies become red and satellite galaxies undergo environmental
quenching. \citet{Bundy2010}, \citet{Ilbert2010}, and
\citet{Pozzetti2010} compare results obtained for galaxies classified
from rest-frame colours and morphology, finding that the
transformation from blue to red colours and from disk-dominated to
bulge-dominated morphologies may be due to two or more processes,
which are either environmentally driven (strangulation, major or minor
merging with varying amounts of gas) or internal (instabilities, gas
consumption, morphological quenching, AGN feedback) \citep{Bundy2010}.
Any scenario should account for the non-negligible fraction of
quiescent disk-dominated galaxies at low masses, and involve processes
with different timescales for the shutdown of the star formation and
the morphological transformation \citep[e.g.][]{Pozzetti2010}, whereas
for massive galaxies the correspondence of red colours and elliptical
morphologies should be explained by a single dominant mechanism,
probably associated with secular evolution \citep{Oesch2010}.  We
explore in more detail the differences between morphological and
colour transformation in different environments in
Sect.~\ref{timescale}.

\citet{Scodeggio2009} study the rest-frame colours of VVDS galaxies 
at $0.2<z<1.4$ in environments based on the density contrast on scales
of $\sim 8$\,Mpc, and conclude that the segregation of galaxy
properties is ultimately the result of the large scale environment,
via the mass of the dark matter halo. This conclusion agrees with our
findings: from Fig.~\ref{fig:mfenv}, we infer that the large-scale
environment sets up the stellar mass distribution, which is in turn
is linked to the mass of the hosting haloes, and its evolution.

At low redshift, the bimodality of the GSMF has also been detected:
for instance, from the SDSS dataset, \citet{Baldry2006} and
\citet{Baldry2008} detect a significant upturn at low stellar masses
with respect to the single Schechter function on the global and
environment dependent GSMFs.

Considering the alternative definition of environment, i.e., galaxy
clusters and groups, we also find in the literature signs of an excess
of low mass systems, for instance by converting the composite LF of
RASS-SDSS clusters by \citet{Popesso2006} to GSMFs by making an
assumption about the mass-to-light ratio, as done in
\citet{Baldry2008}.  A steep low mass end is seen for clusters,
steeper than the upturn noticed in the field from the SDSS and also,
to a lesser extent, than our $\alpha_1$ value in D4 in the low
redshift bin.  The mechanisms responsible for the bimodal nature of
the GSMFs should therefore operate in both the field and high density
environments, but in the most dense regions they should be able to
originate the steepest low mass end.  For instance, in
\citet{Rudnick2009}, the same bimodality in the LF can be seen for
SDSS clusters at low redshifts, and in \citet{Banados2010} for
galaxies members of Abell 1689 at $z=0.183$.  Analyses of high
redshift clusters
\citep[e.g.][]{Poggianti1999,Poggianti2009,Desai2007,SanchezB2009,Simard2009,Patel2009b,Patel2009a,Just2010,Wolf2009,Gallazzi2009,Balogh2007,Balogh2009,Wilman2009,Treu2003}
are mainly focused on the buildup of the red sequence and the
evolution of the fraction of morphological types, in particular S0
galaxies, linked especially to the peculiar mechanisms acting on these
densest environments.

In these quoted works, a complex picture, but broadly consistent
within the uncertainties, is emerging for the evolutionary paths of
galaxies, with many mechanisms playing a role, whose relative
importance is a function of the mass, environment and past history of
each considered system.


\subsection{The mechanism and timescale of galaxy transformation}
\label{timescale}

Figures~\ref{fig:frac_mf_delta_type} and \ref{fig:mcrosst} provide
some clues about the timescale and mechanism responsible of galaxy
quenching in different environments.  We have found that the evolution
in the high density regions is more rapid than in low density ones,
i.e., the rate of transformation into photometric early-types is
higher from $z=1$ to low redshifts in overdense regions than
underdense ones.  Therefore, some of the mechanisms responsible for
quenching the star formation, and then transforming blue galaxies into
passive ones, must be environment dependent.  The physical processes
operating on galaxies and transforming their colours and/or
morphologies can be internally or externally driven and
gravitationally or hydrodynamically induced \citep[for reviews
see][]{Boselli2006,Treu2003}.  Since only a small fraction of the
galaxies studied are probably located in rich clusters, we have not
sought to consider processes that occur primarily in such very high
density environments.  Improbable processes are ram pressure
stripping, consisting of the gas stripping of a galaxy moving through
a dense inter-galactic medium and the abrupt truncation of its star
formation, and harassment, i.e. a gravitational interaction in high
velocity encounters of galaxies, causing morphological transformation
and bursts of star formation.  Given the typical galaxy velocities and
inter-galactic medium density involved in these processes, they cannot
have a significant impact on the results presented in this
paper. Post-starburst galaxies have been found in a wide range of
environments in DEEP2 \citep{Yan2009} and zCOSMOS \citep{Vergani2010}
indicating that the formation mechanism behind this class of objects,
i.e. their star formation shutdown, is not a peculiarity of clusters.

Viable mechanisms in the field are galaxy-galaxy merging and
starvation.  Major merging processes can trigger AGN activity and
quench the star formation: the fraction of pairs, related to the rate
of merging, may depend on environment.  Merging of galaxies in the
densest regions is impeded by the high relative velocities, but at
high redshift, supposedly $z \sim 1$, this process was more common,
thus the merging rate higher \citep{deRavel2009}.  In this context, at
high redshift merging processes produced a shift in the GSMF towards
higher masses, because of the depletion at low masses and consequent
increase in early-type galaxies at high masses.  At later times, the
decrease in the merging rate ensures that the high mass end remains
almost constant, while the acquisition of new galaxies from the field,
by means of the hierarchical growth of the structures, can produce the
observed shape of the D4 GSMF at low redshift in Figs.~\ref{fig:mfenv}
and \ref{fig:mf_delta_type}, the dip at intermediate masses, and
the high contribution of massive early-type galaxies.

To explain the evolution in the density of massive elliptical
galaxies, \citet{Ilbert2010} concluded that the rate of wet mergers
should steeply decline at $z<1$.  Limits on the contribution of major
merging as primary mechanism can be drawn from the evolution of pair
fraction \citep[][who found that 20\% of the stellar mass in present
day galaxies with $\log\mathcal{M}/\mathcal{M}_\odot>9.5$ has been
accreted by major merging events since $z\sim 1$]{deRavel2009} and
from the GSMF \citep[][who derived an average number of total mergers
$\sim 0.16$\,gal$^{-1}$\,Gyr$^{-1}$ since $z\sim 1$ for the global
population, derived from the GSMF evolved according to the mass growth
due to star formation]{Pozzetti2010}.

In addition, strangulation (also referred to starvation or
suffocation), consisting of halo-gas stripping, can play a role: when
the diffuse warm and hot gas reservoir in the galaxy corona is
stripped because of gravitational interaction with low-mass group-size
haloes or with cluster haloes at large distances from the core, the
gas cannot be accreted anymore and the galaxy will exhaust the
remaining cold gas through star formation, on a timescale which can be
instantaneous or slow, i.e., up to a few Gyr, depending on the mass of
the galaxy \citep{Wolf2009}. The result is the suppression of the star
formation, not immediately followed by a morphological transformation,
explaining the possible presence of red spirals, even if the fading of
the disc can lead to an earlier-type morphological classification.
This mechanism alone is not able to reproduce the shape of the D4 GSMF
and the contribution of the different galaxy types, since it predict a
large amount of red galaxies at low masses \citep[for the difficulties
of the starvation scenario see also][]{Bundy2010}, as demonstrated by
comparing observed data with simulations in Sect.~\ref{mock};
nonetheless, this mechanism may be effective in the group environment,
where galaxies are undergoing morphological transformations and
suppression of their star formation \citep[e.g.][]{Wilman2009}.

\begin{figure}
\centering
\includegraphics[width=\hsize]{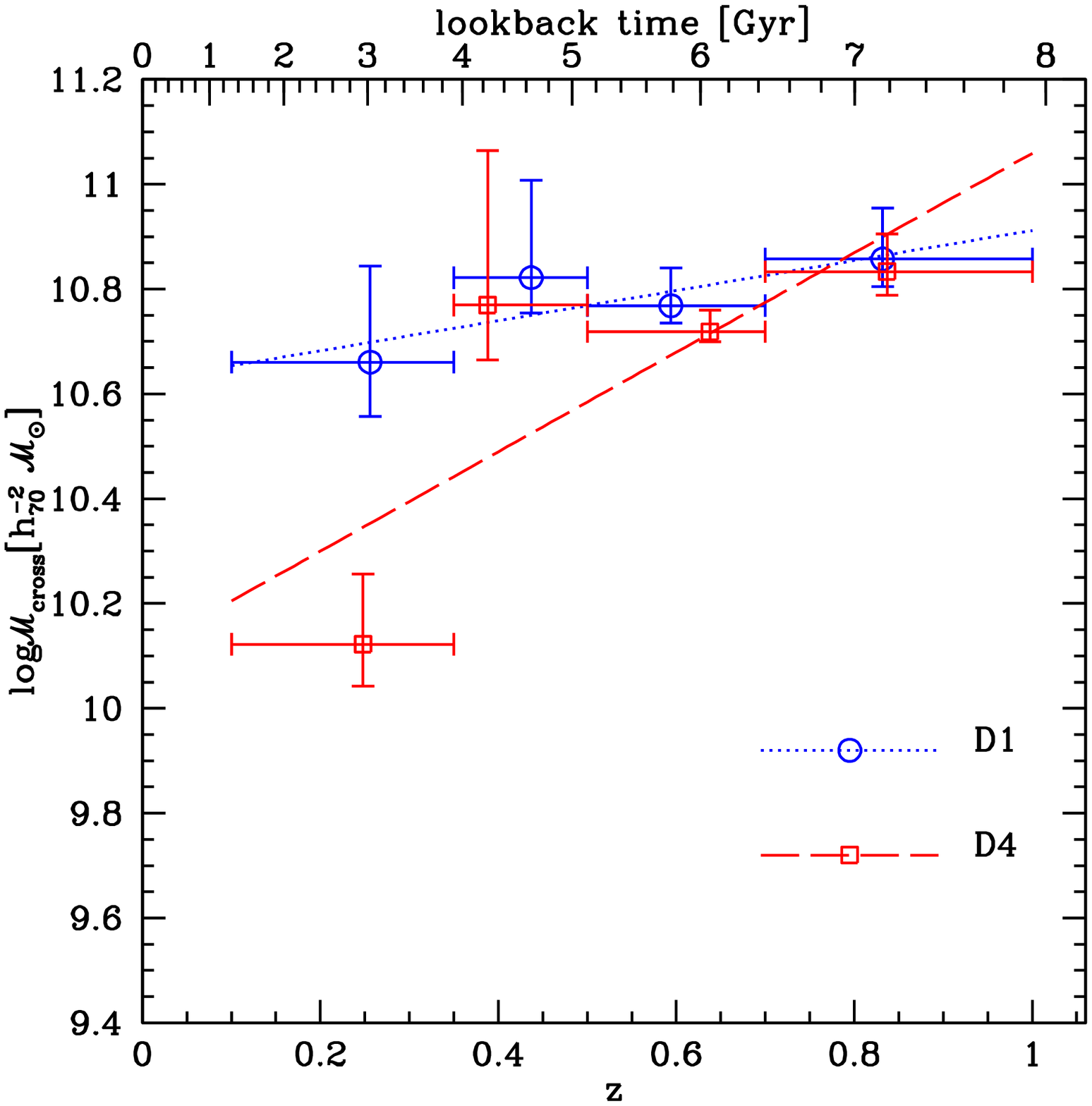}
\caption{
Like Fig.~\ref{fig:mcrosst}, with ${\mathcal M}_{\rm cross}$ computed
for morphological types.}
\label{fig:mcrossm}
\end{figure}

To help identify the most likely transformation mechanisms, we also
computed GSMFs for samples divided following the morphological
classification by \citet{Scarlata2007}, as defined in
Sect.~\ref{class}.  In Fig.~\ref{fig:mcrossm}, we show the values of
${\mathcal M}_{\rm cross}$ in the $4$ considered redshift bins.  This
plot appears to differ from the analogous plot obtained for samples
produced by dividing galaxies on the basis of photometric types: the
values of ${\mathcal M}_{\rm cross}$ are higher and their evolution
seems insensitive to the environment from $z\sim 1$ to $z\sim
0.4$. The higher values of ${\mathcal M}_{\rm cross}$ for the
morphological classification suggest that the dynamical transformation
into elliptical galaxies follows the quenching of their star
formation. It is possible that the transformations of morphology occur
on longer timescales than those of colour
\citep[e.g.]{Capak2007b,Smith2005,Bamford2009,Wolf2009}, as inferred
also from the study of post-starburst galaxies selected in the same
zCOSMOS sample (Vergani et al. \citeyear{Vergani2010}) or by
considering different evolutionary paths \citep{Skibba2009}. A more
comprehensive study should be performed to investigate this point,
since the larger number of photometric early-types than morphological
ones may also be caused by a relatively large fraction of
dust-reddened spiral galaxies.

To evaluate the uncertainties related to this comparison of
photometric and morphological types, we altered the threshold between
elliptical galaxies and morphological late-types: we divided the
morphological class $2.1$, which should still represent
bulge-dominated galaxies, following the observed $B-z$: the
evolutionary track of the $B-z$ colour of a galaxy Sab
\citep{Coleman1980} provides a criterion to separate quiescent and
star-forming galaxies in good agreement with the spectral
classification, as shown in \citet{Mignoli2009}. With this separation,
the values of the morphological ${\mathcal M}_{\rm cross}$ become
consistent with the photometric values, both in terms of the absolute
value and the trend with redshift.

Both mechanisms, gas stripping and interactions, likely operate to
explain the suppression of the star formation and the morphological
transformation. Those processes act on different timescales and have
different efficiencies as a function of galaxy mass and environment,
but it is still difficult to draw firm conclusions, because of the
uncertainties associated with the galaxy classification.


\subsection{Comparison with mock catalogues}
\label{mock}

\begin{figure}
\centering
\includegraphics[width=\hsize]{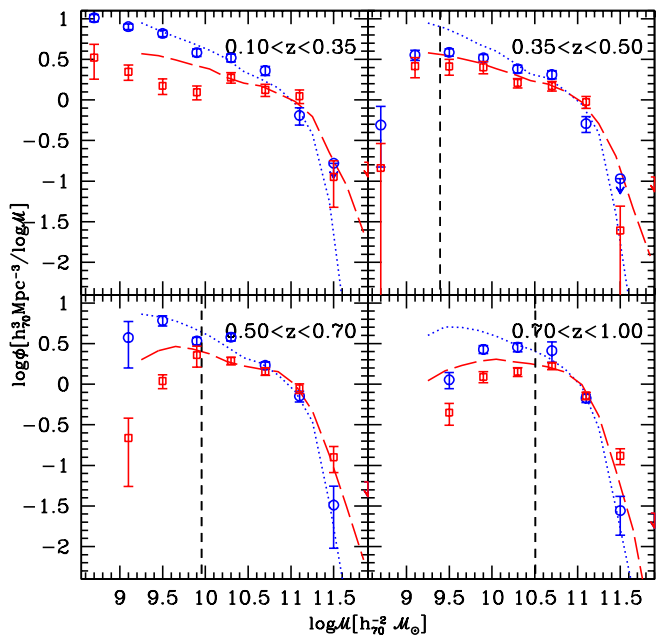}
\caption{
GSMFs derived with $1/V_{\rm max}$ method in mock catalogues (D1
environment: blue dotted lines, D4: red dashed lines, both
representing the average obtained from $12$ mocks) compared to the
observed ones (points) in D1 and D4 environments (blue circles and red
squares, respectively).  The functions are rescaled to arbitrary
units, to maintain the same integral of the GSMFs in the overdense
regions at masses larger than $10^{10.5}\,\mathcal{M}_\odot$ in
observed and mocks samples.}
\label{fig:mf_mock_env}
\end{figure}

\begin{figure*}
\centering
\includegraphics[width=0.48\hsize]{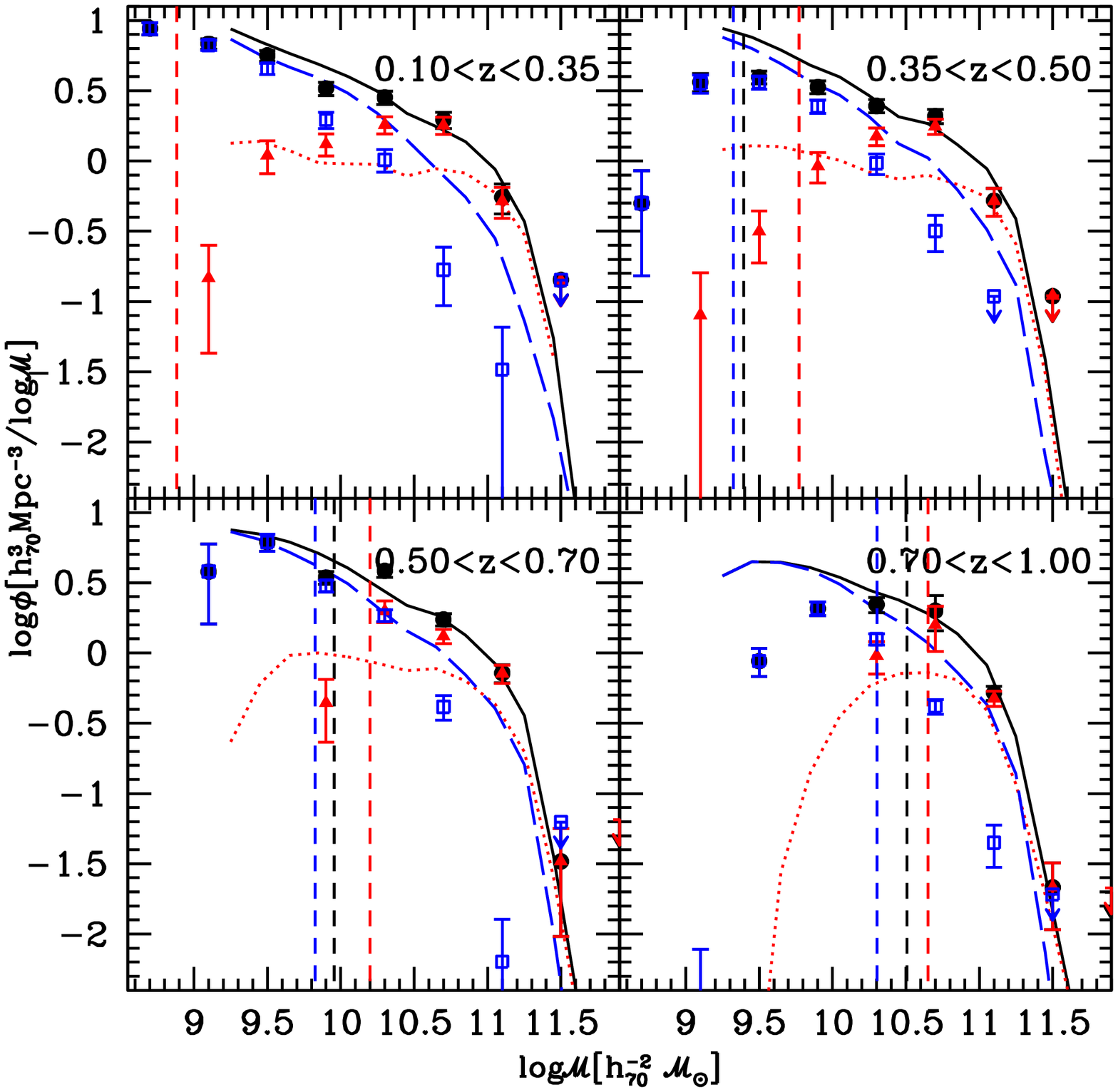}
\includegraphics[width=0.48\hsize]{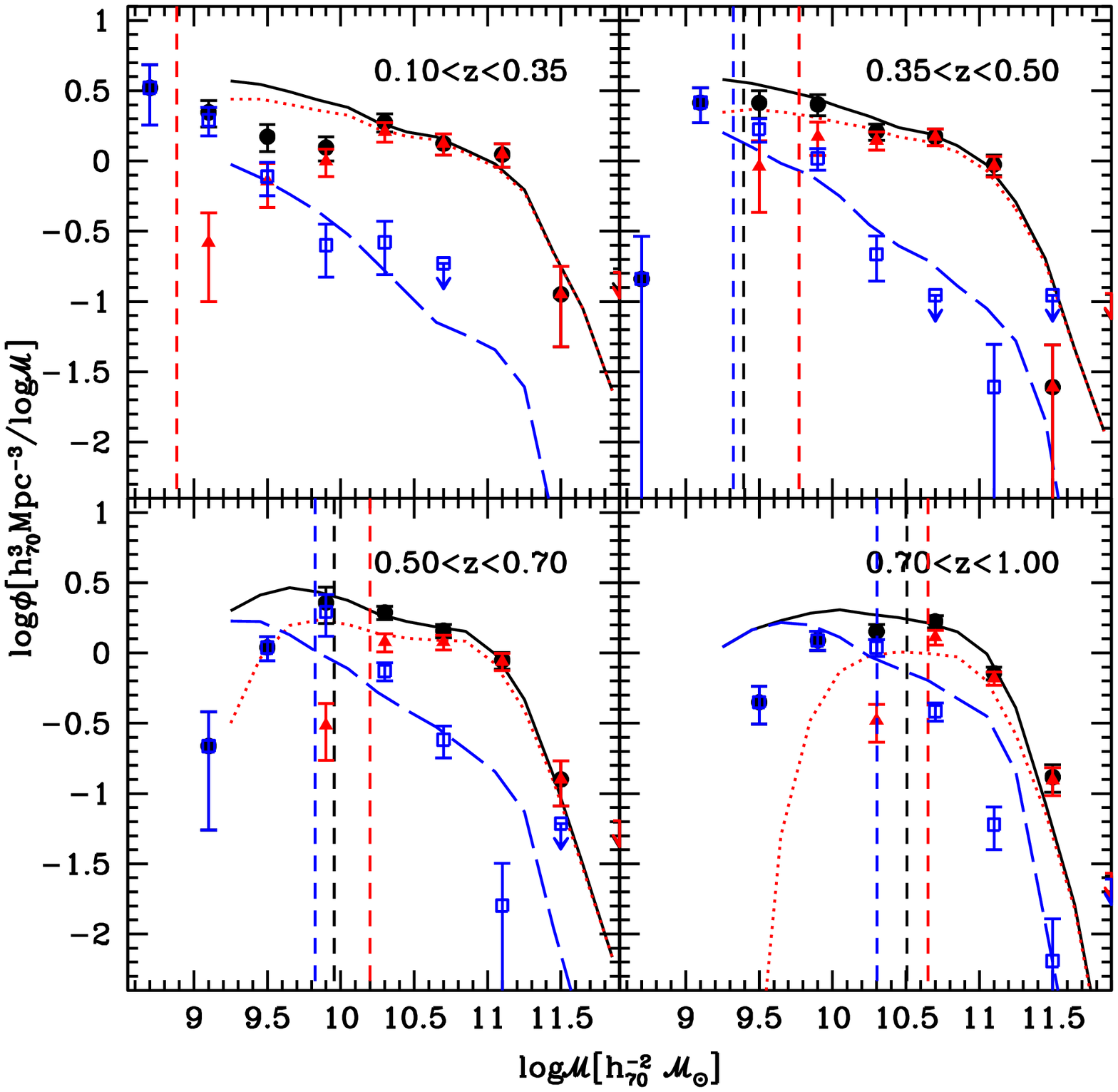}
\caption{
Left: quartile D1 (low density environment). Right: quartile D4 (high
density).  Points refer to the observed quantities, lines to the GSMFs
derived from the mock catalogues. Black points and solid lines: GSMFs
relative to the considered density quartile, renormalised to the same
integral at $\log \mathcal{M}/\mathcal{M}_\odot>10.5$. Red triangles
and dotted lines: galaxies with $B-I>1.15$. Blue squares and
dashed lines: galaxies with $B-I\le1.15$. }
\label{fig:mf_mock_env_typ}
\end{figure*}

We used $12$ COSMOS mock lightcones \citep{Kitzbichler2007} based on
the Millennium N-body simulation \citep{Springel2005}. The galaxy
population of lightcones was then assigned by means of semi-analytical
recipes \citep{Croton2006,DeLucia2007}. The final catalogues are the
same as those described in \citet{Knobel2009}, who used them to test
the group finder algorithm.

We used the 5NN flux-limited $1+\delta$ estimate of the environment
and the rest-frame colour $B-I$ to differentiate early- from late-type
galaxies, and to be able to compare the same quantities in
observations and mocks.  Even though at the lowest stellar masses the
mock catalogues may be affected by colour incompleteness, this does
not affect our analysis, since we limit our comparison to the higher
masses probed in the zCOSMOS.  In Fig.~\ref{fig:mf_mock_env}, we
compare the high and low-density GSMFs in both the observed sample and
the $12$ averaged mock catalogues.  To avoid normalisation
uncertainties caused by cosmic variance \citep{Meneux2009}, we decided
to renormalise the GSMFs, in such a way that the observed and mock
GSMFs of the overdense regions have the same integral value at masses
higher than $10^{10.5}\,\mathcal{M}_\odot$ in all the redshift bins.
The most evident characteristic of the observed GSMFs, namely the
bimodality of the GSMFs in overdense regions at low redshift, is not
reproduced by semi-analytical models.  To explore the reason for this
failure of semi-analytical models (SAMs) in reproducing observations,
we separated red and blue galaxies adopting the threshold $B-I=1.15$,
which corresponds to the location of the dip of the colour bimodality,
obtaining the GSMFs in Fig.~\ref{fig:mf_mock_env_typ}.  For the low
density environments, SAMs produce too many blue galaxies at
intermediate and especially at high masses in all the redshift bins,
and consequently also a too low density of red galaxies, in particular
at $10^{10}-10^{11}\,\mathcal{M}_\odot$. This can be ascribed to an
inefficient suppression of the star formation in the absence of
external drivers, as in the case of sparse environments.
\citet{Weinmann2006} also find a too high blue fraction of central
galaxies: they explain this discrepancy by an improper modelling of
dust extinction, which is very likely underestimated for starburst
galaxies, and AGN feedback, that may be more effective above a given
halo mass. A threshold halo mass above which the star formation is
naturally shut down, as proposed by \citet{Cattaneo2008}, may also
alleviate the discrepancy.

In the high density regions, the most visible difference is the excess
of low and intermediate mass red galaxies
($<10^{10}\,\mathcal{M}_\odot$) in SAMs with respect to the observed
fractions in the lowest redshift bin, where the probed mass range is
wider.  This last comparison reflects the problem of the overquenching
of satellites in the SAMs we used, which produces too many small red
galaxies: a too efficient strangulation produces an instantaneous shut
down of the star formation when a galaxy enters in a halo
\citep[see][for a description of the problem and some attempts to solve
it]{Weinmann2006,Weinmann2010,Font2008,Kang2008,Kimm2009,Fontanot2009}.

%

\section{Conclusions}

We have computed GSMFs in different environments and studied the
relative contributions of different galaxy types to these GSMFs, and
their evolution. Our main results are:

\begin{enumerate}

\item 
The bimodality seen in the global GSMF \citep{Pozzetti2010} up to
$z\sim 0.5$ is considerably more pronounced in high density
environments; a sum of two Schechter functions is thus required to
reproduce the observed non-parametric estimates of the GSMF.

\item 
The bimodality is due to the different relative contributions of
early- and late-type galaxies in different environments, each
contribution being reasonably well represented by a single Schechter
function.

\item 
The shapes of the GSMFs of different galaxy types in different
environments and their evolution with time are very similar, i.e., the
differences on the global GSMFs may be ascribed to the evolution in
the normalisation of the GSMFs of different galaxy types in the
extreme environments we have considered.

\item 
The evolution with time in the fractional contributions of different
galaxy types to the environmental GSMF appears to be a function of the
overdensity in which the galaxies live, and is consistent with a
higher rate of downsizing with time in overdense regions.

\item 
The evolution of the crossover mass for photometric late- and
early-type galaxies suggests a faster transition rate in overdense
regions, with galaxies in low-density regions experiencing the same
evolutionary path as the analogous galaxies in overdense environments
with a delay of $\sim 2$\,Gyr being accumulated between $z\sim 1$ and
$z\sim 0.2$.

\item 
The environment starts to play a significant role in the evolution of
galaxies at $z\la 1$.

\item 
The timescales for quenching of star formation and morphological
metamorphosis differ in different environments; tentatively, the
crossover mass considering morphological classification suggests that
the morphological transformation is slower than the colour change.

\item 
SAMs fail in different ways as a function of the environment: GSMFs
computed from mock catalogues show an underestimate of the number of
red massive galaxies in low density environments, probably because of
an inefficient internal mechanism suppressing the star formation at
relatively high masses; in high density regimes the overquenching
problem of satellites in SAMs causes an excess of red galaxies at
intermediate and low masses.

\end{enumerate}

As a consequence of the remarkable difference in the shape of the
GSMFs in under- and overdense regions, we can infer that all the galaxy
properties depending on mass will also depend on environment by virtue
of the GSMF environmental dependence, as shown in the case of the
colour--density and morphology--density relations
\citep{Cucciati2010,Tasca2009} and of the AGN fraction
\citep{Silverman2009}.

The nature versus nurture debate is unresolvable, because the mass of
a galaxy, often thought to be its nature, is a strong function of the
environment. A more relevant issue is the understanding of the
mechanisms producing the observed evolution of galaxies and their
transition from late- to early-type in different environments.

Future investigations will also concern the impact of merging in
different environments (\citealp{deRavel2010}; Kampczyk et
al. \citeyear{Kampczyk2010}) and the role of the dark-matter halo mass
functions in different environments \citep[e.g.][]{Abbas2007} in the
determining galaxy formation efficiency.

\begin{acknowledgements}
MB wishes to thank Preethi Nair, Alexis Finoguenov, and Ramin Skibba
for useful discussions, comments and suggestions.  We thank the
anonymous referee for the constructive first report, that helped
improve the paper. MB is grateful to the editor, Fran\c{c}oise Combes,
for her kind support.

This work was partly supported by an INAF contract
PRIN/2007/1.06.10.08 and an ASI grant ASI/COFIS/WP3110 I/026/07/0.

\end{acknowledgements}

\bibliographystyle{aa}
\bibliography{bolzonella_zcosmos_v2}

\end{document}